\documentclass[a4paper]{article}
\usepackage{geometry}
\usepackage{setspace}
\usepackage{float}
\usepackage{epstopdf}
\usepackage{standalone}
\usepackage{amsmath}
\usepackage{pdflscape}
\usepackage{authblk}
\usepackage{graphicx}
\usepackage[flushleft]{threeparttable}
\usepackage[comma,authoryear]{natbib}
\usepackage[citecolor=blue]{hyperref}

\newcommand{\interior}[1]{%
 {\kern0pt#1}^{\mathrm{o}}%
}
\usepackage [english]{babel}
\usepackage [autostyle, english = american]{csquotes}
\MakeOuterQuote{"}

\usepackage{algorithm,algpseudocode}
\usepackage{caption}
\newcommand{\Tau}{\mathrm{T}}
\usepackage[makeroom]{cancel}

\title{\textbf{Bayesian Inference in the Presence of Intractable Normalizing Functions}}
\author[1]{Jaewoo Park}
\author[1]{Murali Haran}
\affil[1]{Department of Statistics, The Pennsylvania State University}

\begin{document}

\maketitle

\begin{abstract}

Models with intractable normalizing functions arise frequently in statistics. Common examples of such models include exponential random graph models for social networks and Markov point processes for ecology and disease modeling. Inference for these models is complicated because the normalizing functions of their probability distributions include the parameters of interest. In Bayesian analysis they result in so-called doubly intractable posterior distributions which pose significant computational challenges. Several Monte Carlo methods have emerged in recent years to address Bayesian inference for such models. We provide a framework for understanding the algorithms and elucidate connections among them. Through multiple simulated and real data examples, we compare and contrast the computational and statistical efficiency of these algorithms and discuss their theoretical bases. Our study provides practical recommendations for practitioners along with directions for future research for MCMC methodologists.

\end{abstract}

\noindent%


\vfill

\section{Introduction}

~~~~~Markov chain Monte Carlo (MCMC) has been used to routinely carry out Bayesian inference for an enormous variety of complicated models \citep[cf.][]{brooks2011handbook}. However, inference for Bayesian models with intractable normalizing functions, where the normalizing constant of the model is itself a function of parameters of interest, is still far from routine. There are many well known models that have intractable functions, for instance the class of exponential family random graph models and its variants \citep[cf.][]{robins2007introduction, hunter2012inference} which are popular models for social networks. Non-Gaussian Markov random field models in spatial statistics \citep[cf.][for a review]{besag1974spatial, hughes2011autologistic} also have intractable normalizing functions. In fact, it is worth noting that the Ising model \citep{lenz1920beitrag,ising1925beitrag}, which is a non-Gaussian Markov random field, appears in the landmark paper on the Metropolis algorithm \citep{metropolis1953equation}. While the Metropolis algorithm provides an elegant way to simulate from this model for a given parameter value, the paper did not consider the more difficult problem of performing inference for this model.

Consider $h(\mathbf{x}|\theta)$, an unnormalized probability model for a random variable $\mathbf{x} \in \mathcal{X}$ given a parameter vector $\theta \in \Theta$, with a normalizing function $Z(\theta)=\int_{\mathcal{X}} h(\mathbf{x}|\theta)d\mathbf{x}$. Let $p(\theta)$ be the prior density for $\theta$. The likelihood function, $L(\theta|\mathbf{x})$ is $h(\mathbf{x}|\theta)/Z(\theta)$ and the posterior density of $\theta$ is

\begin{equation}
\pi(\theta|\mathbf{x}) \propto p(\theta)\frac{h(\mathbf{x}|\theta)}{Z(\theta)}.
\label{e1}
\end{equation}

The problem in constructing an MCMC algorithm stems from the fact that $Z(\theta)$ cannot be easily evaluated and the acceptance probability at each step of the Metropolis-Hastings (MH) algorithm \citep{metropolis1953equation,hastings1970monte} involves evaluating $Z(\theta)$ both at the current and proposed value of $\theta$.

There have been several proposals to address the intractable normalizing function problem in a maximum likelihood context. \cite{besag1974spatial} proposed the maximum pseudolikelihood estimate (MPLE) which maximizes the pseudolikelihood, a particular likelihood approximation that does not require $Z(\theta)$. MPLE does not, in general, define a likelihood function. Furthermore, MPLE may be a reasonable estimator when there is weak dependence among data points relative to the size of the data set but in other cases its performance is unsatisfactory. \cite{younes1988estimation} proposed a stochastic gradient algorithm to solve normal equations to find the maximum likelihood estimate (MLE) for models with intractable normalizing functions. However the step size and starting point must be selected carefully, otherwise the algorithm becomes slow and does not converge \citep{ibanez2003parameter}. \cite{geyer1992constrained} propose MCMC-MLE which is based on maximizing a Monte Carlo approximation to the likelihood; this approximation is based on an importance sampling approximation of $Z(\theta)$. This is an elegant and theoretically justified algorithm. In practice, MCMC-MLE suffers from some of the usual challenges faced by importance sampling approaches, namely that for an accurate approximation to MLE, the initial value for the algorithm should be reasonably close to the MLE. Some of these issues may be partially addressed by  umbrella sampling \citep{torrie1977nonphysical,geye:2011}, which involves sampling from mixtures of importance sampling distributions. However approximating standard errors can also be difficult in situations where analytical gradients for the unnormalized likelihood are unavailable and using bootstrap techniques may be computationally infeasible \citep[cf.][]{goldstein2014attraction}. In such cases  and also in situations where there is an interest in incorporating prior information about the parameters or avoiding model degeneracies \citep[cf.][]{handcock2002statistical}, Bayesian alternatives may be preferable. In this manuscript, we focus on Bayesian inference; we refer readers to \cite{geye:2011} for a general overview and \cite{hunter2012computational} for a review of recent MCMC-MLE methods for exponential random graph models.

Several MCMC algorithms have recently been proposed for Bayesian inference in the presence of intractable normalizing functions. These algorithms may be broadly classified into two general if somewhat overlapping categories: (1) algorithms where the introduction of a well chosen auxiliary variable results in the normalizing function (or a ratio of normalizing functions) canceling out in the Metropolis-Hastings acceptance probability, and (2) directly approximating the normalizing function (or a ratio of normalizing functions), and substituting the approximation into the acceptance probability. Here we will refer to the first as an {\it auxiliary variable approach} and the second as a {\it likelihood approximation approach} \citep[see also a discussion in][]{liang2015adaptive}. In addition, MCMC algorithms may also be classified as “asymptotically exact” or “asymptotically inexact”. For asymptotically exact algorithms the Markov chain's stationary distribution is exactly equal to the desired posterior distribution. On the other hand, asymptotically inexact or "noisy" algorithms generate Markov chains without this property; even asymptotically the samples generated only follow the target distribution approximately.

In what follows we discuss several MCMC algorithms. We provide an explanation of the ideas underpinning each algorithm along with figures that summarize them and make it easier to see how they are related. We also discuss theoretical justifications and practical implementation issues. We carry out a comparative study by using three different examples: an Ising model, a social network model, and a spatial point process. We provide some guidance about potential advantages and disadvantages of each algorithm along with connections among them, providing some future avenues for research. The remainder of this paper is organized as follows. In Section 2 we discuss several auxiliary variable algorithms. In Section 3 we cover several likelihood approximation algorithms. In Section 4 we describe the application of the algorithms in the context of three different case studies and provide some insights based on our results. In particular, we discuss in detail the computational complexity of the algorithms in Section 5. We point out connections between the algorithms in Section 6 along with general guidelines and recommendations based on our study. We conclude with a summary in Section 7.

\section{Auxiliary Variable Approaches}

~~~~~In this section, we review several auxiliary variable approaches. Here, the target distribution includes both the parameter of interest as well as an auxiliary variable. By a clever choice of the auxiliary variable proposal the intractable functions get cancelled in the acceptance probability of the Metropolis-Hastings algorithm. What distinguishes the different auxiliary variable algorithms from each other is how the auxiliary variable is sampled.

\subsection{Auxiliary Variable MCMC}

~~~~~\cite{moller2006efficient} introduce an auxiliary variable $\mathbf{y}$ with the conditional density $f(\mathbf{y}|\theta,\mathbf{x})$ so that the intractable terms are cancelled in the acceptance probability of the Metropolis-Hastings algorithm. Suppose the original target density is $\pi(\theta|\mathbf{x}) \propto p(\theta)h(\mathbf{x}|\theta)/Z(\theta)$. Then the augmented target density is $\pi(\theta,\mathbf{y}|\mathbf{x}) \propto f(\mathbf{y}|\theta,\mathbf{x})p(\theta)h(\mathbf{x}|\theta)/Z(\theta)$ whose marginal density becomes $\int_{\mathcal{X}} \pi(\theta,\mathbf{y}|\mathbf{x}) d\mathbf{y} = \pi(\theta|\mathbf{x})$,  the original target density. Now consider a joint proposal density for $\lbrace \theta,\mathbf{y} \rbrace$ updates which can be factorized as $q(\theta',\mathbf{y}'|\theta,\mathbf{y})=q(\mathbf{y}'|\theta')q(\theta'|\theta)$. \cite{moller2006efficient} take the proposal for the auxiliary variable $q(\mathbf{y}'|\theta')$ as $h(\mathbf{y}'|\theta')/Z(\theta')$ which is the same as the model for the data $\mathbf{x}$ given the parameter value $\theta'$ proposed from $q(\theta'|\theta)$. The resulting algorithm for the above joint density with this proposal density has Metropolis-Hastings acceptance probability

\begin{equation}
   \alpha = \min\left\lbrace 1, \frac{f(\mathbf{y}'|\theta',\mathbf{x})p(\theta')h(\mathbf{x}|\theta')\bcancel{Z(\theta)}h(\mathbf{y}|\theta)\bcancel{Z(\theta')}q(\theta|\theta')}{f(\mathbf{y}|\theta,\mathbf{x})p(\theta)h(\mathbf{x}|\theta)\bcancel{Z(\theta')}h(\mathbf{y}'|\theta')\bcancel{Z(\theta)}q(\theta'|\theta)}\right\rbrace.
\label{e3}
\end{equation}

\begin{figure}
\begin{center}
\includegraphics[scale=1.4]{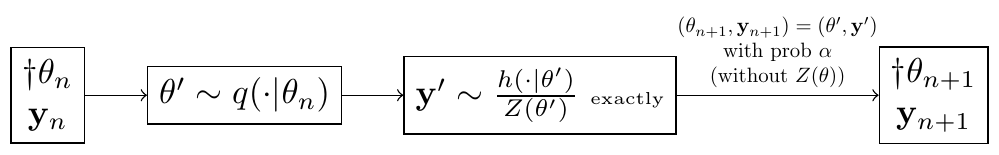}
\end{center}
\caption[]{Illustration of the (n+1) step update for auxiliary variable MCMC. The dagger symbols indicate the parameter of interest.}
\label{AVMFig}
\end{figure}

The resulting Metropolis-Hastings acceptance probability \eqref{e3} does not contain the normalizing functions because they get cancelled out. Since $\int_{\mathcal{X}} \pi(\theta,\mathbf{y}|\mathbf{x}) d\mathbf{y} = \pi(\theta|\mathbf{x})$, the marginal $\theta$ samples follow the original target distribution. We will henceforth use AVM to refer to this auxiliary variable MCMC algorithm. We summarize the algorithm in Figure~\ref{AVMFig}. We should note that updating $\mathbf{y}$ requires drawing a sample with exact distribution $h(\cdot|\theta')/Z(\theta')$. For some models this can be achieved via perfect sampling \citep{propp1996exact}, a clever method that uses bounding Markov chains to construct a sampler where the draws are exactly (not just asymptotically) from the target distribution. Although perfect sampling is possible for some models, for instance certain Markov random field (MRF) models, it is not trivial to construct a perfect sampler in general. This is a major practical limitation of the AVM.

\begin{algorithm}
\caption{Auxiliary variable MCMC (AVM)}\label{AVMalg}
\begin{algorithmic}[H]
\normalsize
\State Given $\lbrace \theta_{n},\mathbf{y}_{n} \rbrace \in \Theta \times \mathcal{X}$ at $n$th iteration.
\State 1. Propose $\theta' \sim q(\cdot|\theta_{n})$.
\State 2. Propose the auxiliary variable exactly from probability model at $\theta'$: $\mathbf{y}' \sim \frac{h(\cdot|\theta')}{Z(\theta')}$ using perfect sampling.
\State 3. Accept $\lbrace \theta_{n+1},\mathbf{y}_{n+1}\rbrace=\lbrace \theta',\mathbf{y}'\rbrace$ with probability

$\alpha=\min\left\lbrace 1, \frac{f(\mathbf{y}'|\theta',\mathbf{x})p(\theta')h(\mathbf{x}|\theta')h(\mathbf{y}_{n}|\theta_{n})q(\theta_{n}|\theta')}{f(\mathbf{y}_{n}|\theta_{n},\mathbf{x})p(\theta_{n})h(\mathbf{x}|\theta_{n})h(\mathbf{y}'|\theta')q(\theta'|\theta_{n})}\right\rbrace$, else reject (set $\lbrace \theta_{n+1},\mathbf{y}_{n+1}\rbrace=\lbrace \theta_n, \mathbf{y}_n \rbrace$).

\end{algorithmic}
\end{algorithm}

\textbf{Components to be tuned}: \cite{moller2006efficient} report that the choice of $f(\mathbf{y}|\theta,\mathbf{x})$ impacts the mixing of the chain. Suppose $Z(\theta)$ is known so that we can choose $f(\mathbf{y}|\theta,\mathbf{x})=h(\mathbf{y}|\theta)/Z(\theta)$. Then it is easily seen that  \eqref{e3} is identical to the acceptance probability of the Metropolis-Hastings algorithm with the stationary density $\pi(\theta|\mathbf{x})$, which implies that this chain has the same convergence properties as the Markov chain where the normalizing function is known and the same proposal $q(\theta'|\theta)$ is used for $\theta$. Therefore, a reasonable choice of conditional density for the auxiliary variable is $f(\mathbf{y}|\theta,\mathbf{x})$ that approximates $h(\mathbf{y}|\theta)/Z(\theta)$. A simple choice is $f(y|\theta,\mathbf{x})=h(\mathbf{y}|\widehat{\theta})/Z(\widehat{\theta})$, where $\widehat{\theta}$ may be an approximation to the MLE. $\widehat{\theta}$ should be predetermined before implementing the AVM algorithm. For example, the maximum pseudolikelihood estimate (MPLE) proposed by \cite{besag1974spatial} may be an option though for some problems the MPLE may be a poor approximation to the MLE. Then $f(\mathbf{y}'|\theta',\mathbf{x})/f(\mathbf{y}|\theta,\mathbf{x})=h(\mathbf{y}'|\widehat{\theta})/h(\mathbf{y}|\widehat{\theta})$, which makes it possible to calculate \eqref{e3} because there are no intractable terms. Another possible choice of $f(\mathbf{y}|\theta,\mathbf{x})$ is using a normalizable density without $Z(\theta)$. AVM mixes better as $f(\mathbf{y}|\theta,\mathbf{x})$ more closely resembles $h(\mathbf{y}|\theta)/Z(\theta)$.

\textbf{Theoretical justification}: The Markov chain satisfies the detailed balance condition if $\pi(\theta|\mathbf{x})T(\theta'|\theta) = \pi(\theta'|\mathbf{x})T(\theta|\theta')$, where
$T(\theta'|\theta)$ is the transition kernel of the Markov chain with the target density $\pi(\theta|\mathbf{x})$. Since the Markov Chain satisfies detailed balance with respect to the augmented target distribution, the marginal distribution of which is the posterior distribution of $\theta$, AVM is an asymptotically exact MCMC algorithm. However, to achieve the detailed balance condition, we need to sample $\mathbf{y}$ exactly from the likelihood function $h(\mathbf{y}|\theta')/Z(\theta')$.

\subsection{The Exchange Algorithm}

~~~~~Appearing almost simultaneously with \cite{moller2006efficient}, the exchange algorithm \citep{murray2006} also constructs an augmented target distribution and updates the augmented state via the Metropolis-Hastings algorithm. Consider the auxiliary variable $\mathbf{y}$ which follows $h(\mathbf{y}|\theta')/Z(\theta')$ and conditional density of $\theta'$ for given $\theta$ as $q(\theta'|\theta)$  where $\theta$ is the parameter setting of data $\mathbf{x}$. Then the augmented joint density is

\begin{equation}
   \pi(\theta,\theta',\mathbf{y}|\mathbf{x}) \propto p(\theta)L(\theta|\mathbf{x})q(\theta'|\theta)L(\theta'|\mathbf{y})=p(\theta)\frac{h(\mathbf{x}|\theta)}{Z(\theta)}q(\theta'|\theta)\frac{h(\mathbf{y}|\theta')}{Z(\theta')}.
\label{e4}
\end{equation}

\begin{figure}
\begin{center}
\includegraphics[scale=1.4]{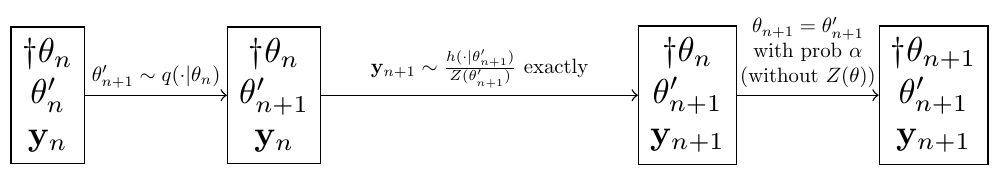}
\end{center}
\caption[]{Illustration for the exchange algorithm. The dagger symbols indicate the parameter of interest.}
\label{ExchangeFig}
\end{figure}

For this augmented density, $\lbrace \theta',\mathbf{y} \rbrace$ is updated through block-Gibbs samplers; $\theta'$ is generated from the proposal $q(\cdot|\theta)$ and the auxiliary variable $\mathbf{y}$ is generated from $h(\cdot|\theta')/Z(\theta')$. The first two arrows in Figure~\ref{ExchangeFig} correspond to the update of $\lbrace \theta',\mathbf{y} \rbrace$. Then $\theta$ is updated through exchanging parameter settings. Let $\lbrace \theta,\theta' \rbrace$ be the current parameter settings for $\lbrace \mathbf{x},\mathbf{y} \rbrace$. Consider a swapping proposal $s(\lbrace \theta^{*},\theta'^{*} \rbrace | \lbrace \theta,\theta' \rbrace) = \delta(\theta^{*}-\theta')\delta(\theta'^{*}-\theta)$, where $\delta$ denotes the Dirac delta function. After swapping is proposed, data $\mathbf{x}$ follows $\theta'$ instead of $\theta$ and the auxiliary variable $\mathbf{y}$ follows $\theta$ instead of $\theta'$. The symmetric swapping proposal results in the acceptance probability ($\alpha$),

\begin{equation}
   \min\left\lbrace 1, \frac{\bcancel{
s(\lbrace \theta,\theta' \rbrace|\lbrace \theta^{*},\theta'^{*} \rbrace)
}\pi(\theta',\theta,\mathbf{y}|\mathbf{x})}{\bcancel{s(\lbrace \theta^{*},\theta'^{*} \rbrace | \lbrace \theta,\theta' \rbrace)
}\pi(\theta,\theta',\mathbf{y}|\mathbf{x})}\right\rbrace = \min\left\lbrace 1, \frac{p(\theta')h(\mathbf{x}|\theta')\bcancel{Z(\theta)}h(\mathbf{y}|\theta)\bcancel{Z(\theta')}q(\theta|\theta')}{p(\theta)h(\mathbf{x}|\theta)\bcancel{Z(\theta')}h(\mathbf{y}|\theta')\bcancel{Z(\theta)}q(\theta'|\theta)}\right\rbrace,
\label{e5}
\end{equation}

with intractable terms cancelled. Since $\int_{\mathcal{X}} \int_{\Theta} \pi(\theta, \theta',\mathbf{y}|\mathbf{x}) d\theta'd\mathbf{y} = \pi(\theta|\mathbf{x})$, the marginal $\theta$ samples follow the original target distribution, as shown in Figure~\ref{ExchangeFig}. This idea of swapping parameter settings is connected to the parallel tempering algorithm \citep{geyer1991markov}. Parallel tempering swaps the states while preserving joint distributions in the product space. The exchange algorithm also swaps parameter settings between $\mathbf{x}$ and $\mathbf{y}$, while preserving the augmented distribution. Since we do not need to keep $\lbrace \theta',\mathbf{y} \rbrace$ in the actual implementation of the algorithm, the exchange algorithm may be simply written as Algorithm~\ref{EXalg}.

\begin{algorithm}
\caption{Exchange algorithm }\label{EXalg}
\begin{algorithmic}[H]
\normalsize
\State Given $\theta_{n} \in \Theta$ at $n$th iteration.
\State 1. Propose $\theta' \sim q(\cdot|\theta_{n})$.
\State 2. Generate the auxiliary variable exactly from probability model at $\theta'$: $\mathbf{y} \sim \frac{h(\cdot|\theta')}{Z(\theta')}$ using perfect sampling.
\State 3. Accept $\theta_{n+1}=\theta'$ with probability

$\alpha=\min\left\lbrace 1, \frac{p(\theta')h(\mathbf{x}|\theta')h(\mathbf{y}|\theta_{n})q(\theta_{n}|\theta')}{p(\theta_{n})h(\mathbf{x}|\theta_{n})h(\mathbf{y}|\theta')q(\theta'|\theta_{n})}\right\rbrace$, else reject (set $\lbrace \theta_{n+1},\mathbf{y}_{n+1}\rbrace=\lbrace \theta_n, \mathbf{y}_n \rbrace$).
\end{algorithmic}
\end{algorithm}

The exchange algorithm can also be extended through "bridging" \citep{murray2006}, Algorithm~\ref{EXbridgingalg}. The idea may be summarized as follows. In the acceptance probability of Algorithm~\ref{EXalg}, one may think of the ratio $h(\mathbf{x}|\theta')/h(\mathbf{x}|\theta)$ as measuring whether the proposed $\theta'$ explains the data $\mathbf{x}$ better than current $\theta$ or not. Also, $h(\mathbf{y}|\theta)/h(\mathbf{y}|\theta')$ represents how well the auxiliary variable $\mathbf{y}$ is supported under $\theta$ versus $\theta'$. Therefore, even if $\theta'$ is a much better candidate than $\theta$ under the data $\mathbf{x}$, swapping can be rejected if the auxiliary variable $\mathbf{y}$ is improbable under $\theta$, which can lead to slow mixing of the chain. To improve this, annealed importance sampling (AIS) \citep{neal1996sampling,neal2001annealed} can be combined with the exchange algorithm. Instead of sampling a single auxiliary variable, a series of auxiliary variables $\lbrace \mathbf{y}_{0},\mathbf{y}_{1},...,\mathbf{y}_{K} \rbrace$ are sampled from the intermediate densities between $h(\mathbf{y}|\theta')$ and $h(\mathbf{y}|\theta)$. Intermediate densities can be constructed as

\begin{equation}
f_{k}(\mathbf{y}|\theta,\theta') = h(\mathbf{y}|\theta')^{1-\beta_{k}}h(\mathbf{y}|\theta)^{\beta_{k}},\beta_{k}=\frac{k}{K+1},~k=1, \dots ,K.
\label{e6}
\end{equation}

After proposing $\theta'$, generate $\mathbf{y}_{0}$ from $h(\mathbf{y}|\theta')/Z(\theta')$. Then generate each $\mathbf{y}_{k}$ from $T_{k}(\mathbf{y}_{k}|\mathbf{y}_{k-1})$, the Metropolis-Hastings (MH) transition kernel whose stationary density is $f_{k}(\mathbf{y}|\theta,\theta')$. This bridging step helps to generate $\mathbf{y}_{K}$ that is more probable under $\theta$. 

\begin{algorithm}
\caption{Exchange algorithm with bridging }\label{EXbridgingalg}
\begin{algorithmic}[H]
\normalsize
\State Given $\theta_{n} \in \Theta$ at $n$th iteration.
\State 1. Propose $\theta' \sim q(\cdot|\theta_{n})$.

\State 2.  Generate the series of auxiliary variables $\lbrace \mathbf{y}_{0},\mathbf{y}_{1},...,\mathbf{y}_{K} \rbrace$ from \eqref{e6}:

$\mathbf{y}_{0} \sim \frac{h(\cdot|\theta')}{Z(\theta')}$ using perfect sampling.

$\mathbf{y}_{k} \sim T_{k}(\mathbf{y}_{k}|\mathbf{y}_{k-1})$ via 1-step MH update for $k=1,...,K$.

\State 3. Accept $\theta_{n+1}=\theta'$ with bridging probability

$\alpha=\min\left\lbrace 1, \frac{p(\theta')h(\mathbf{x}|\theta')q(\theta_{n}|\theta')}{p(\theta_{n})h(\mathbf{x}|\theta_{n})q(\theta'|\theta_{n})}\prod_{k=0}^{K}\frac{f_{k+1}(\mathbf{y}_{k}|\theta_{n},\theta')}{f_{k}(\mathbf{y}_{k}|\theta_{n},\theta')}\right\rbrace$,

else reject (set $\lbrace \theta_{n+1},\mathbf{y}_{n+1}\rbrace=\lbrace \theta_n, \mathbf{y}_n \rbrace$).
\end{algorithmic}
\end{algorithm}

\textbf{Components to be tuned}: The basic exchange algorithm does not require any tuning besides the usual tuning of the proposal for $\theta$. For the extended version, there are many options for bridging but even here \eqref{e6} provides an automated schedule. Larger $K$ in \eqref{e6} can lead to better mixing of the chain at the expense of computing time. Effective sample size (ESS) \citep{kass1998markov} provides a rough diagnostic for how well the chain is mixing. Therefore, effective sample size per unit time (ESS/T) can be used to determine $K$ that provides a compromise between mixing and computational cost \citep{murray2007advances}.

\textbf{Theoretical justification}: Both the exchange algorithm and the extended exchange algorithm with bridging are asymptotically exact MCMC in that constructed Markov chains satisfy detailed balance condition with respect to the target distribution. However just like AVM, the exchange algorithm is of limited applicability because it requires perfect sampling to generate the auxiliary variable.

\subsection{The Double Metropolis-Hastings Sampler}

\begin{figure}
\begin{center}
\includegraphics[scale=1.4]{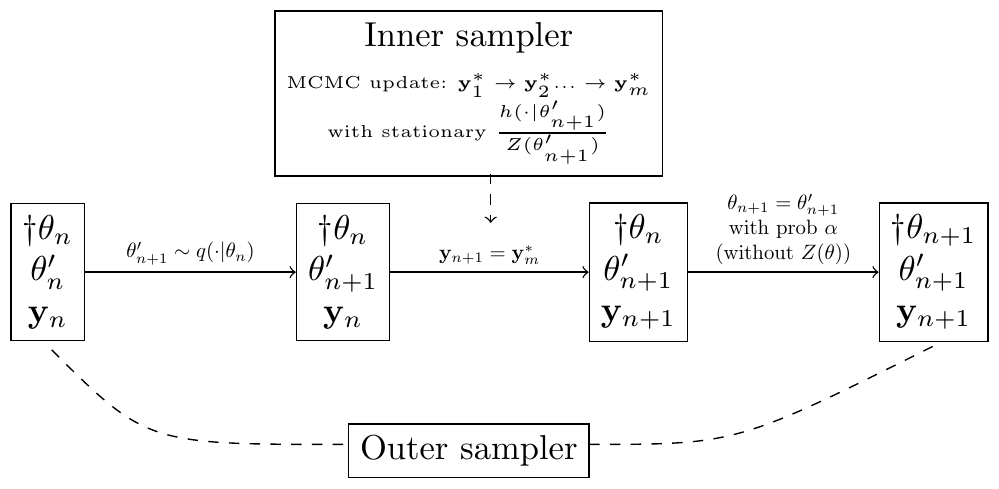}
\end{center}
\caption[]{Illustration of the Double Metropolis-Hastings algorithm. The dagger symbols indicate the parameter of interest.}
\label{DMHFig}
\end{figure}

~~~~~Both AVM and the exchange algorithm require perfect sampling which can be very expensive or impossible for complicated probability models. \cite{liang2010double} replaces perfect sampling for the auxiliary variable with a standard Metropolis-Hastings algorithm; the last state of the resulting Markov chain is then treated like a draw from the perfect sampler. This is called Double Metropolis-Hastings (DMH) because a Metropolis-Hastings algorithm is used within another Metropolis-Hastings algorithm. As in Figure~\ref{DMHFig}, one is the "outer sampler" to generate $\theta$ draws while the other is the "inner sampler" used to generate the auxiliary variable $\mathbf{y}$.

Define $T^{m}_{\theta'}(\mathbf{y}|\mathbf{x})$ as $m$-MH updates from $\mathbf{x}$ to $\mathbf{y}$ under $\theta'$ whose stationary density is $h(\mathbf{y}|\theta')/Z(\theta')$; this satisfies the detailed balanced condition

\begin{equation}
   h(\mathbf{x}|\theta')T^{m}_{\theta'}(\mathbf{y}|\mathbf{x}) = h(\mathbf{y}|\theta')T^{m}_{\theta'}(\mathbf{x}|\mathbf{y}).
\label{e7}
\end{equation}

Plugging this result into \eqref{e5} yields

\begin{equation}
   \alpha = \min\left\lbrace 1, \frac{p(\theta')h(\mathbf{y}|\theta)T^{m}_{\theta'}(\mathbf{x}|\mathbf{y})q(\theta|\theta')}{p(\theta)h(\mathbf{x}|\theta)T^{m}_{\theta'}(\mathbf{y}|\mathbf{x})q(\theta'|\theta)}\right\rbrace,
\label{e8}
\end{equation}

which is the acceptance probability of the DMH algorithm. DMH approximates perfect sampling through $m$ steps of an (inner) Metropolis-Hastings algorithm. A similar approximate approach which replaces perfect sampling with MH updates for the social network models is discussed in \cite{caimo2011bayesian}. As we show later, DMH is simpler to implement and computationally more efficient relative to all other algorithms. However, its computational efficiency directly depends on the efficiency of the inner sampler. If the inner sampler update $T^{m}_{\theta'}(\mathbf{y}|\mathbf{x})$ is expensive or needs large $m$ to get probable auxiliary variable $\mathbf{y}$, then DMH becomes computationally expensive. Since $m$ is usually proportional to the dimension of the data $\mathbf{x}$, the inner sampler is the main computational bottleneck for large data.

\begin{algorithm}
\caption{Double Metropolis-Hastings algorithm }\label{DMHalg}
\begin{algorithmic}[H]
\normalsize
\State Given $\theta_{n} \in \Theta$ at $n$th iteration.
\State 1. Propose $\theta' \sim q(\cdot|\theta_{n})$.
\State 2. Generate the auxiliary variable approximately from probability model at $\theta'$:

$\mathbf{y} \sim T^{m}_{\theta'}(\cdot|\mathbf{x})$ using $m$-MH updates.

\State 3. Accept $\theta_{n+1}=\theta'$ with probability

$\alpha = \min\left\lbrace 1, \frac{p(\theta')h(\mathbf{x}|\theta')h(\mathbf{y}|\theta_{n})q(\theta_{n}|\theta')}{p(\theta_{n})h(\mathbf{x}|\theta_{n})h(\mathbf{y}|\theta')q(\theta'|\theta_{n})}\right\rbrace = \min\left\lbrace 1, \frac{p(\theta')h(\mathbf{y}|\theta_{n})T^{m}_{\theta'}(\mathbf{x}|\mathbf{y})q(\theta_{n}|\theta')}{p(\theta_{n})h(\mathbf{x}|\theta_{n})T^{m}_{\theta'}(\mathbf{y}|\mathbf{x})q(\theta'|\theta_{n})}\right\rbrace$,

else reject (set $\lbrace \theta_{n+1},\mathbf{y}_{n+1}\rbrace=\lbrace \theta_n, \mathbf{y}_n \rbrace$).
\end{algorithmic}
\end{algorithm}

\textbf{Components to be tuned}: Users need to decide the number of MH updates $m$ to generate $\mathbf{y}$. There are numerous stopping rules for MCMC if computing expectations is the goal \citep[cf.][]{flegal2008markov,gong2015practical}. However, here determining convergence to the stationary distribution is of interest \citep{jones2001honest, rosenthal1995minorization}, which is a very challenging problem in general. A simple heuristic for determining $m$ is to set $m$ to be proportional to the size of the data, for instance $m=5n$, where $n$ is the size of the data. We recommend that DMH should be run for larger $m$ values, for instance $m=10n$ to confirm that the results do not change much as $m$ is increased.

\textbf{Theoretical justification}: DMH is an asymptotically inexact algorithm because the detailed balance condition does not hold for the outer sampler unless the inner sampler length ($m$) approaches infinity; in practice the inner sampler length is of course finite.

\subsection{Adaptive Exchange Algorithm}

\begin{figure}
\begin{center}
\includegraphics[scale=1.4]{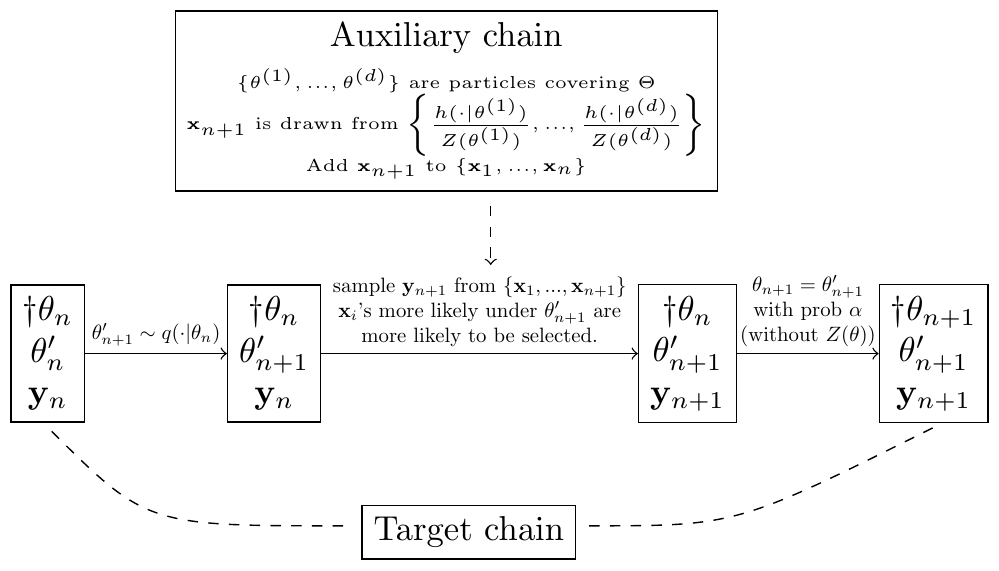}
\end{center}
\caption[]{Illustration for the Adaptive Exchange algorithm. The dagger symbols indicate the parameter of interest.}
\label{AEXFig}
\end{figure}

~~~~~In order to address the asymptotic inexactness of DMH, \cite{liang2015adaptive} propose the adaptive exchange algorithm (AEX). AEX runs two chains simultaneously: the auxiliary chain and the target chain (see Figure~\ref{AEXFig}). The auxiliary chain generates a sample from a family of distributions, $\lbrace h(\mathbf{x}|\theta^{(1)})/Z(\theta^{(1)}),...,h(\mathbf{x}|\theta^{(d)})/Z(\theta^{(d)}) \rbrace$, where $\lbrace \theta^{(1)},...,\theta^{(d)} \rbrace$ are pre-specified "particles" covering a parameter space. The generated sample is stored at each iteration. Then the target chain generates a posterior sample from the $\pi(\theta|\mathbf{x})$ via the exchange algorithm. The difference is that the auxiliary variable $\mathbf{y}$ is sampled from the cumulative samples in the auxiliary chain until current iteration (resampling). With increasing iterations, cumulative samples from the auxiliary chain grow, so that the resampling of $\mathbf{y}$ in the target chain becomes identical to exact sampling of $\mathbf{y}$. Then similar to the exchange algorithm, the marginal density of $\theta$ converges to the target $\pi(\theta|\mathbf{x})$. AEX is therefore an asymptotically exact algorithm that does not require perfect sampling. We begin with some notation for the $n$th iteration of AEX. 

\begin{itemize}
\item Particles: $\lbrace \theta^{(1)},...,\theta^{(d)} \rbrace$, each $\theta^{(i)} \in \Theta$. These particles are fixed through the algorithm.
\item Particle index: $I_{n} \in \lbrace 1,...,d \rbrace$ returns index of a chosen particle at $n$th iteration.
\item Auxiliary data: $\mathbf{x}_{n} \in \mathcal{X}$ is an approximate sample from the probability model at selected particle $\theta^{(I_{n})}$. That is, $\mathbf{x}_{n} \sim h(\cdot|\theta^{(I_{n})})/Z(\theta^{(I_{n})})$ .
\item Normalizing function approximation at each particle: $\mathbf{w}_{n}=\lbrace w^{(1)}_{n},...,w^{(d)}_{n} \rbrace \in W$. For $i=1,...,d$, as $n$ gets large $w^{(i)}_{n}$ converges to $Z(\theta^{(i)})$ via the stochastic approximation algorithm \citep{liang2007stochastic}.
\item Cumulative information: $H_{n} = \cup_{j=1}^{n}\lbrace I_{j},\mathbf{x}_{j},\mathbf{w}_{j} \rbrace$ is necessary for constructing the algorithm. Therefore, $\lbrace I_{n},\mathbf{x}_{n},\mathbf{w}_{n} \rbrace$ should be stored at each iteration.
\item Posterior sample: $\theta_{n} \in \Theta$
\end{itemize}

AEX updates a non-Markovian stochastic process $\lbrace I_{n+1},\mathbf{x}_{n+1},\mathbf{w}_{n+1},\theta_{n+1} \rbrace \in  \lbrace 1,...,d \rbrace \times \mathcal{X} \times R^{d} \times \Theta$ in the $(n+1)$st iteration. For each iteration, the auxiliary chain generates a sample $\mathbf{x}_{n+1}$ from the mixture density, $(1/d)\sum_{i=1}^{d}h(\mathbf{x}|\theta^{(i)})/Z(\theta^{(i)})$ through stochastic approximation Monte Carlo (SAMC) \citep{liang2007stochastic}. Particles are $d$-number of points in the grid over a parameter space $\Theta$. Particles are chosen so that they can cover the important region of the parameter space and the sample spaces of neighboring distributions $\lbrace h(\mathbf{x}|\theta^{(i)})/Z(\theta^{(i)}) \rbrace$ overlap each other. A clever choice of particles can improve mixing of the auxiliary chain and is important for AEX to be a practical algorithm. We refer the reader to the supplementary material for strategies for choosing particles. For the $(n+1)$st update of AEX, cumulative information $H_{n+1}$ is constructed through the auxiliary chain. For $(n+1)$st iteration of the algorithm, $\lbrace I_{n+1},\mathbf{x}_{n+1},\mathbf{w}_{n+1} \rbrace$ is updated for given $\lbrace I_{n},\mathbf{x}_{n},\mathbf{w}_{n} \rbrace$ using SAMC and then added to $H_{n}$ to construct $H_{n+1}$. Hence, the accumulated information $(H_{n+1})$ becomes larger with each iteration of AEX. See Algorithm~\ref{AEXalg1} for details.

Then the target chain generates $\theta_{n+1}$ from the posterior using the exchange algorithm. For proposed $\theta'$, the auxiliary variable $\mathbf{y}$ is sampled from $\lbrace \mathbf{x}_{1},...,\mathbf{x}_{n+1}\rbrace$ through a dynamic importance sampling procedure. The resampling probability is

\begin{equation}
P(\mathbf{y}=\mathbf{x}_{l}|\theta') \propto \sum_{j=1}^{n+1}w^{(I_{j})}_{j}\frac{h(\mathbf{x}_{j}|\theta')}{h(\mathbf{x}_{j}|\theta^{(I_{j})})}1\lbrace \mathbf{x}_{j}=\mathbf{x}_{l}\rbrace,~~l=1,...,n+1.
\label{AEXe1}
\end{equation}

In \eqref{AEXe1}, $h(\mathbf{x}_{j}|\theta^{(I_{j})})/w^{(I_{j})}_{j}$ is the importance function at $j$th iteration. Since the importance function changes as $I_{j}$ varies, it is called a dynamic importance function. Here, $\mathbf{x}_{l}$'s more likely under $\theta'$ are more likely to be sampled. \cite{liang2015adaptive} show that the distribution of resampled $\mathbf{y}$ converges to $h(\mathbf{y}|\theta')/Z(\theta)$ as $n$ increases. Since the proposal for $\mathbf{y}$ in \eqref{AEXe1} changes with each iteration, AEX appears at first to be an adaptive MCMC algorithm. However, unlike basic adaptive MCMC algorithms the target distribution also varies as $H_{n+1}$ grows.

In Step 2(a) of Algorithm~\ref{AEXalg1}, $\lbrace a_{n} \rbrace$ is called the gain factor, which is the step size of update for the abundance factor $\mathbf{w}_{n}$. A large gain factor can lead to faster movement around all $\lbrace h(\mathbf{x}|\theta^{(i)})/Z(\theta^{(i)}) \rbrace$. Since $a_{n}=n_{0}/\max(n_{0},n)$ becomes smaller as $n$ increases, movement between particles becomes slower with increasing iterations. \cite{liang2007stochastic} suggest larger $n_{0}$ for complicated problems so that makes it fast to move around all state space. Step 2(b) illustrates varying truncation \citep{jin2013fitting}, which can help the convergence of $\mathbf{w}_{n}$ regardless of starting point. For $\mathbf{w}_{n} \in W$, consider the sequence of set $\lbrace K_{s} \rbrace \in W$, which satisfies $\cup_{s\geq 0}K_{s}=W$ and $K_{s}\subset K_{s+1}^{\mathrm{o}}$, where $K_{s+1}^{\mathrm{o}}$ denotes the interior of set $K_{s+1}$. A truncation function $\Tau$ maps a point in $\mathcal{X} \times W$ to a random point in $\mathcal{X}_{0} \times K_{0}$ for $\mathcal{X}_{0}\subset \mathcal{X}$, and $\sigma_{n}$ is the number of truncations that occurred by the $n$th iteration ($\sigma_{0}=0$). From such truncation, when an updated $\mathbf{w}_{n+1/2}$ is outside the interesting target region $(K_{s})$, it is reinitialized with a smaller gain factor and expanded target region $(K_{s+1} \supset K_{s})$. This varying truncation can help SAMC to find an appropriate step size $\lbrace a_{n} \rbrace$ and starting point of abundance factor automatically \citep{jin2013fitting}. The reader may understand such varying truncation method as a safeguard for when the abundance factor becomes degenerate.

\begin{algorithm}[hh]
\caption{Adaptive Exchange algorithm \textbf{(Part 1)}: Auxiliary chain\\
Constructing cumulative information $H_{n+1}$}\label{AEXalg1}
\begin{algorithmic}[H]

\normalsize

\State Initialize $\lbrace I_{0},\mathbf{x}_{0},\mathbf{w}_{0} \rbrace$, for example $\mathbf{x}_{0}=$ data $\mathbf{x}$, $I_{0} \sim \lbrace 1,...,d \rbrace$ uniformly, $\mathbf{w}_{0} = \mathbf{1}$. Set $H_{0}=\lbrace I_{0},\mathbf{x}_{0},w^{(I_{0})}_{0} \rbrace$.

\State For $(n+1)$st update, given $ \lbrace I_{n},\mathbf{x}_{n},\mathbf{w}_{n},\theta_{n} \rbrace \in \lbrace 1,...,d \rbrace \times \mathcal{X} \times W \times \Theta$.

\State 1. Update $I_{n+1}$ or $\mathbf{x}_{n+1}$ with equal probability.

\State (a) With probability $0.5$, update $I_{n+1}$ at $\mathbf{x}_{n}$.

Propose $\theta^{(I')}$ from the $k$-nearest particles (say in Euclidean distance) of $\theta^{(I_{n})}$ 

with equal probability.

Accept $\lbrace I_{n+1},\mathbf{x}_{n+1} \rbrace=\lbrace I',\mathbf{x}_{n} \rbrace$ with $\alpha_{1}=\min\left\lbrace 1, \frac{w^{(I_{n})}_{n}h(\mathbf{x}_{n}|\theta^{(I')})}{w^{(I')}_{n}h(\mathbf{x}_{n}|\theta^{(I_{n})})}\right\rbrace$.

\State (b) With probability $0.5$, update $\mathbf{x}_{n+1}$ at $\theta^{(I_{n})}$.

Propose $\mathbf{x}'$ through $m$-MH updates: $\mathbf{x}' \sim T_{I_{n}}^{m}(\cdot|\mathbf{x}_{n})$ whose target is $\frac{h(\mathbf{x}|\theta^{(I_{n})})}{Z(\theta^{(I_{n})})}$.

Accept $\lbrace I_{n+1},\mathbf{x}_{n+1} \rbrace=\lbrace I_{n},\mathbf{x}' \rbrace$ with $\alpha_{2}=\min\left\lbrace 1, \frac{h(\mathbf{x}'|\theta^{(I_{n})})T(\mathbf{x}_{n}|\mathbf{x}')}{h(\mathbf{x}_{n}|\theta^{(I_{n})})T(\mathbf{x}'|\mathbf{x}_{n})}\right\rbrace$.

\State 2. Update approximation of $Z(\theta^{(i)})$ at $i=1,...,d$:

\State Following stochastic approximation \citep{liang2007stochastic}

\State (a) $\log{(w^{(i)}_{n+1/2})}=\log{(w^{(i)}_{n})}+a_{n+1}(~1\lbrace I_{n+1}=i\rbrace-1/d~),~~a_{n}=n_{0}/\max(n_{0},n)$.

\State (b) Stochastic truncation ($\Tau$) is implemented if $\mathbf{w}_{n+1/2}$ is outside of the target region ($K_{\sigma_{n}}$), and records the number of truncation ($\sigma_{n}$):

$\lbrace \mathbf{w}_{n+1},\mathbf{x}_{n+1} \rbrace = \begin{cases}

\Tau(\lbrace \mathbf{w}_{n},\mathbf{x}_{n} \rbrace),~\sigma_{n+1}=\sigma_{n}+1 & \mathbf{w}_{n+1/2} \not\in K_{\sigma_{n}} \\

    \lbrace \mathbf{w}_{n+1/2},\mathbf{x}_{n+1} \rbrace,~\sigma_{n+1}=\sigma_{n}  & o.w.

\end{cases}$

\State 3. Cumulative information is updated: $H_{n+1}=H_{n} \cup \lbrace I_{n+1},\mathbf{x}_{n+1},w^{(I_{n+1})}_{n+1} \rbrace$.

\end{algorithmic}
\end{algorithm}

\clearpage

\begin{algorithm}[hh]
\ContinuedFloat
\caption{Adaptive Exchange algorithm \textbf{(Part 2)}: Target chain\\
Obtain $\theta_{n+1}$ approximately from $\pi(\theta|\mathbf{x})$ by exchange algorithm, using resampled $\mathbf{y}$}\label{AEXalg2}
\begin{algorithmic}[H]

\normalsize
\State Continued from Part 1: for $(n+1)$st update,

\State 4. Propose $\theta'$: $\theta' \sim q(\cdot|\theta_{n})$.

\State 5. Sample the auxiliary variable from collected dataset in auxiliary chain:

$\mathbf{y} \sim \lbrace \mathbf{x}_{1},...,\mathbf{x}_{n+1} \rbrace$, with resampling probabilities as

$P(\mathbf{y}=\mathbf{x}_{l}|\theta') \propto \sum_{j=1}^{|H_{n+1}|}w^{(I_{j})}_{j}\frac{h(\mathbf{x}_{j}|\theta')}{h(\mathbf{x}_{j}|\theta^{(I_{j})})}1\lbrace\mathbf{x}_{j}=\mathbf{x}_{l}\rbrace,~~l=1,...,|H_{n+1}|$.

\State 6. Accept $\theta_{n+1}=\theta'$ with probability

$\alpha_{3}=\min\left\lbrace 1, \frac{p(\theta')h(\mathbf{x}|\theta')h(\mathbf{y}|\theta_{n})q(\theta_{n}|\theta')}{p(\theta_{n})h(\mathbf{x}|\theta_{n})h(\mathbf{y}|\theta')q(\theta'|\theta_{n})}\right\rbrace$, else reject (set $\lbrace \theta_{n+1},\mathbf{y}_{n+1}\rbrace=\lbrace \theta_n, \mathbf{y}_n \rbrace$).

\end{algorithmic}
\end{algorithm}

In practice, \cite{liang2015adaptive} suggest that only the auxiliary chain (Algorithm~\ref{AEXalg1} Part 1) is implemented $N_{1}$ iterations at first to construct the database, and then the auxiliary and target chain can be run simultaneously. In detail, through $N_{1}$ preliminary iterations of the auxiliary chain, information: $\cup_{j=1}^{N_{1}} \lbrace I_{j},\mathbf{x}_{j},\mathbf{w}_{j} \rbrace$ is constructed. Using this information as initial $H_{0}=\cup_{j=1}^{N_{1}} \lbrace I_{j},\mathbf{x}_{j},\mathbf{w}_{j} \rbrace$, we can implement the entire Algorithm~\ref{AEXalg1}. In this case, the size of cumulative information become $|H_{0}|=N_{1}$, and $|H_{n+1}|=N_{1}+n+1$ (without preliminary step, $|H_{0}|=0$, and $|H_{n+1}|=n+1$). The reason of conducting preliminary step is because the early performance of the target chain can be affected by initial starting values of the auxiliary chain. This preliminary step can construct an initial database $H_{0}$ and improves the mixing of the target chain.

Although we classify AEX as an auxiliary variable approach, AEX also shares characteristics with likelihood approximation approaches. The abundance factor $\lbrace w^{(i)}_{j} \rbrace$ converges to $\lbrace Z(\theta^{(i)}) \rbrace$ for each particle, which is guaranteed by SAMC. This implies that likelihood approximation approaches are applied to estimate $Z(\theta^{(i)})$ directly. Then the auxiliary variable is resampled to cancel out $Z(\theta)$ in the target chain. A strength of AEX is its broad applicability, while still remaining asymptotically exact. It is reasonably efficient computationally. However because of its adaptive structure, there can be serious memory issues. When there are low-dimensional sufficient statistics, as is typically the case with exponential family models, only the sufficient statistics of $\mathbf{x}_{n+1}$ need to be stored in Step 3 of the Algorithm~\ref{AEXalg1}. However, in the absence of such sufficient statistics, $\mathbf{x}_{n+1}$ itself need to be stored at each step. Furthermore, without sufficient statistics, resampling probability calculations in Step 5 becomes expensive because $h(\mathbf{x}_{j}|\theta')$ should be recalculated; when there are sufficient statistics, one can simply take the product of $\theta'$ and the sufficient statistic.

\textbf{Components to be tuned}: We provide details about choosing particles (number of particles, strategies for choosing particles) in the supplementary material. In the auxiliary chain, the number of neighbors for updating $\theta^{(I_{n})}$, the number of preliminary runs for the auxiliary chain ($N_{1}$), the number of MH updates ($m$),  gain factor component ($n_{0}$), and the target region ($K_{s}$) of abundance factor should be tuned. For particles, neighbor is defined using the distance from each other. We can define the closest 20 points as neighboring particles, and this choice appears to work well both in simulated examples in this manuscript as well as the examples in \cite{liang2015adaptive}. For updating $\mathbf{x}_{n}$, $m=1$ cycle of MH updates are enough, because we don't need to generate independent samples. For complicated problems \cite{liang2015adaptive} recommend larger values for $N_{1}$ and $n_{0}$. For the simulation examples, \cite{liang2015adaptive} use around $N_{1}=10,000d$, $n_{0}=25,000$ where $d$ is the number of particles. We used similar values in our simulated examples. If the preliminary auxiliary chain does not appear to converge, larger $n_{0}$ and $N_{1}$ should be used. The convergence of the auxiliary chain can be checked by comparing sampling frequencies $v(i)/N_{1}$ with the target probabilities $1/d$, where $v(i)$ is the number of visitations at each particle. If $v(i)/N_{1} \approx 1/d$ for each particle, then the preliminary run of auxiliary chain can be diagnosed as having converged. For the choice of compact subset of $W$, \cite{liang2015adaptive} set $K_{s}=[0,10^{100+10s}]^{d}$ and $\mathcal{X}_{0}=\mathcal{X}$ in their examples. Detailed simulation settings about the social network model also can be checked in \cite{jin2013bayesian}. 

\textbf{Theoretical justification}: The AEX is an asymptotically exact algorithm and the ergodic average satisfies the Weak Law of Large Numbers. Since AEX is adaptive in both proposal and target, conventional theory for adaptive MCMC does not directly apply. \cite{liang2015adaptive} extend a result in \cite{roberts2007coupling}[Theorem 1], which shows the ergodicity of the adaptive chain with the same target distribution. Although the original proof \citep{liang2015adaptive} assumes compactness of the parameter space $\Theta$, \cite{jin2013bayesian} extend the results without a compactness assumption for $\Theta$. The details of the assumptions for the proof(s) are provided across several results in \cite{jin2013bayesian}[Lemma 3.1, 3.2 and Theorem 3.1]. In order to make it easy for readers to understand the assumptions from a practical point of view, we have attempted to distill the key assumptions as follows: (1) both $\mathcal{X}$ and $W$ are compact, (2) $h(\mathbf{x}|\theta)$ is bounded away from $0$ and $\infty$, (3) $\lim_{n\rightarrow \infty}a_{n}=0, ~ \sum_{n=1}^{\infty}a_{n}=\infty, ~ \sum_{n=1}^{\infty}a^{\eta}_{n}<\infty$ for some $\eta \in (1,2]$, (4) Doeblin condition holds for kernel of auxiliary chain. Some of these conditions are easily verified in practice. The sample space $\mathcal{X}$ for data is compact for finite lattice problem and point process with finite number of points on bounded domain. Sample space $W$ for abundance factor is also compact, if we set $K_{s}$ as suggested in the components to be tuned above. The second assumption is also satisfied for realistic parameter settings. The third is a technical assumption for SAMC, and is satisfied if we construct $a_{n}=n_{0}/\max(n_{0},n)$. The sufficient condition for the last assumption is local positiveness of $m$-MH updates in Step 1(b) of Algorithm~\ref{AEXalg1}, which means $\exists ~ \delta >0, \epsilon >0$ such that, $\forall \mathbf{x} \in \mathcal{X}, |\mathbf{x}-\mathbf{y}|\leq \delta$ implies $T(\mathbf{y}|\mathbf{x})\geq \epsilon$. This is satisfied for a simple Gibbs sampler, birth-death sampler in point process, and tie-no-tie sampler in social network models. Therefore, broadly speaking, the assumptions hold in most problems. We point the readers to the manuscript in order to find the complete set of assumptions for these results. 

\section{Likelihood Approximation Methods}

~~~~While the auxiliary variable approaches we have discussed so far avoid approximating $Z(\theta)$, the approaches based on likelihood approximations we discuss in this section directly approximate $Z(\theta)$ through Monte Carlo and substitute the approximation into the Metropolis-Hastings acceptance probability.

\subsection{Atchade, Lartillot and Robert's Adaptive MCMC}

~~~~~\cite{atchade2008bayesian} provide an adaptive MCMC algorithm, henceforth the ALR algorithm, which approximates $Z(\theta)$ through importance sampling in the acceptance probability. ALR constructs a non-Markovian stochastic process and approximates $Z(\theta)$ at every iteration using the entire sample path of the process. ALR is an asymptotically exact algorithm that does not require perfect sampling.

The ALR algorithm utilizes importance sampling ideas developed in MCMC-MLE \citep{geyer1992constrained}. Consider $\lbrace \mathbf{x}_{1},...,\mathbf{x}_{n} \rbrace$, samples from $h(\mathbf{x}|\theta^{(0)})/Z(\theta^{(0)})$ for some $\theta^{(0)}$. Then $Z(\theta)/Z(\theta^{(0)})$ is an expectation that can be approximated by importance sampling. However the approximation might be poor if $\theta$ is far from $\theta^{(0)}$. \cite{atchade2008bayesian} introduces multiple particles, $\lbrace \theta^{(1)},...,\theta^{(d)} \rbrace$, and a linear combination of importance sampling estimates for $\lbrace Z(\theta)/Z(\theta^{(1)}),...,Z(\theta)/Z(\theta^{(d)}) \rbrace$ approximates $Z(\theta)$. Larger weights are assigned to the estimate of $Z(\theta)/Z(\theta^{(i)})$ when $\theta$ and $\theta^{(i)}$ are closer to each other. This idea of sampling from a mixture is related to umbrella sampling \citep{torrie1977nonphysical} which provides a more robust approximation than a single point importance sampling estimate, as long as one of the $\theta^{(i)}$s is close to $\theta$. We begin with some notation for the $n$th iteration of ALR. 

\begin{figure}
\begin{center}
\includegraphics[scale=1.8]{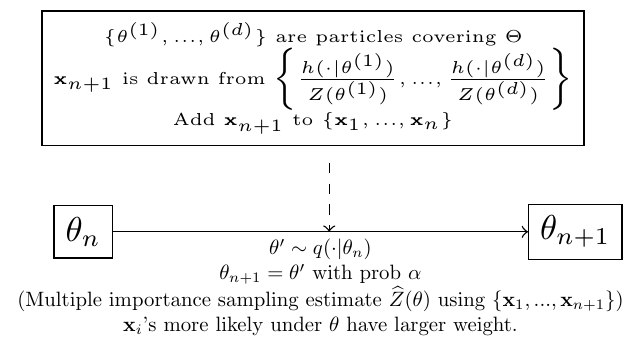}
\end{center}
\caption[]{Illustration for Atchade, Lartillot and Robert's (ALR) algorithm.}
\label{ALRFig}
\end{figure}

\begin{itemize}
\item Particles: $\lbrace \theta^{(1)},...,\theta^{(d)} \rbrace$, each $\theta^{(i)} \in \Theta$. These particles are fixed through the algorithm.
\item Particle index: $I_{n} \in \lbrace 1,...,d \rbrace$ returns the index of the selected particle at $n$th iteration.
\item Auxiliary data: $\mathbf{x}_{n} \in \mathcal{X}$ is an approximate sample from the probability model at a selected particle $\theta^{(I_{n})}$. i.e. $\mathbf{x}_{n} \sim h(\cdot|\theta^{(I_{n})})/Z(\theta^{(I_{n})})$ .
\item Normalizing function approximation at each particle: $\mathbf{c}_{n}=\lbrace c^{(1)}_{n},...,c^{(d)}_{n} \rbrace \in  R^{d}$. For $i=1,...,d$, as $n$ gets large $c^{(i)}_{n}$ converges to $\log{Z(\theta^{(i)})}$ by stochastic approximation \citep{wang2001efficient,atchade2004wang}.
\item Cumulative information: $H_{n} = \cup_{j=1}^{n}\lbrace I_{j},\mathbf{x}_{j} \rbrace$ is necessary for constructing the algorithm. Therefore, $\lbrace I_{n},\mathbf{x}_{n} \rbrace$ should be stored at each iteration.
\item Posterior sample: $\theta_{n} \in \Theta$.
\end{itemize}

By extending a result in \cite{wang2001efficient,atchade2004wang}, the ALR algorithm constructs a non-Markovian stochastic process $\lbrace \mathbf{x}_{n+1},I_{n+1},\mathbf{c}_{n+1},\theta_{n+1} \rbrace \in \mathcal{X} \times \lbrace 1,...,d \rbrace \times R^{d} \times \Theta$. $Z(\theta)$ is approximated through multiple importance sampling and plugged into the acceptance probability to generate samples from $\pi(\theta|\mathbf{x})$. A stochastic approximation for estimating  marginal densities, similar to the approximation used in ALR, is also described in  \cite{liang2007continuous}. The decomposition of $Z(\theta)$ is

\begin{equation}
\begin{split}
    Z(\theta) & = \int_{\mathcal{X}} h(\mathbf{x}|\theta) d\mathbf{x} \\
    & = \sum_{i=1}^{d}k(\theta,\theta^{(i)})\int_{\mathcal{X}} \frac{h(\mathbf{x}|\theta)}{h(\mathbf{x}|\theta^{(i)})}h(\mathbf{x}|\theta^{(i)})d\mathbf{x} \\
    & = \sum_{i=1}^{d}k(\theta,\theta^{(i)})Z(\theta^{(i)})\int_{\mathcal{X}} \frac{h(\mathbf{x}|\theta)}{h(\mathbf{x}|\theta^{(i)})}\frac{h(\mathbf{x}|\theta^{(i)})}{Z(\theta^{(i)})}d\mathbf{x}. \\
\label{e9}
\end{split}
\end{equation}

This suggests the following Monte Carlo approximation of $Z(\theta)$,

\begin{equation}
\widehat{Z}_{n+1}(\theta) = \sum_{i=1}^{d}k(\theta,\theta^{(i)})\exp(c_{n+1}^{(i)})\frac{\sum_{k=1}^{n+1}\frac{h(\mathbf{x}_{k}|\theta)}{h(\mathbf{x}_{k}|\theta^{(i)})}1\lbrace I_{k}=i\rbrace}{\sum_{k=1}^{n+1}1\lbrace I_{k}=i \rbrace},
\label{e10}
\end{equation}


 a weighted sum of the importance sampling estimate $Z(\theta^{(i)})$ at each particle. A "similarity kernel" $k(\theta,\theta^{(i)})$ measures the distance between $\theta$ and $\theta^{(i)}$ and assigns larger weights to particles closer to $\theta$. Similar to AEX, particles are pre-specified before starting the algorithm and are chosen so that they can cover the parameter space. In the examples in \cite{atchade2008bayesian}, particles are collected from the stochastic approximation \citep{younes1988estimation}, though particles could also be chosen as in AEX (see supplementary material for details). Algorithm~\ref{Atchadealgp1} provides ALR details.

\begin{algorithm}
\caption{ALR algorithm \textbf{(Part 1)}: \\
Find good initial values $\lbrace \mathbf{x}_{\tau},I_{\tau},\mathbf{c}_{\tau} \rbrace$ for use in Part 2}\label{Atchadealgp1}

\begin{algorithmic}[H]
\normalsize

\State Initialize $\lbrace \mathbf{x}_{0},I_{0},\mathbf{c}_{0} \rbrace$, for example $\mathbf{x}_{0}=$ data $\mathbf{x}$, $I_{0}\sim \lbrace 1,...,d\rbrace$ uniformly, $\mathbf{c}_{0}=\mathbf{0}$.

\State For $(n+1)$st update, given $\lbrace \mathbf{x}_{n},I_{n},\mathbf{c}_{n} \rbrace \in \mathcal{X} \times \lbrace 1,...,d \rbrace \times R^{d}$.

\While{$\gamma_{n+1} > \epsilon_{1}$} ($\gamma_{n}$ is step size used in step 3 below)\\

Reset $\mathbf{v}=\mathbf{0} \in R^{d}$ ($v(i)$ counts the number of visitations at each particle $\theta^{(i)}$)

\While{$\max_{i}|v(i)-\bar{v}| > \epsilon_{2}\bar{v}$} (until each particle has been visited equally)\\

~~~~ For $i=1,...,d$, set $v(i)=v(i)+1_{i}(I_{n+1})$.\\

~~~~ 1. Update $\mathbf{x}_{n+1}$ through $m$-MH step for given particle:
~~~~ \State $\mathbf{x}_{n+1} \sim T_{I_{n}}^{m}(\cdot|\mathbf{x}_{n})$ whose stationary density is $\frac{h(\mathbf{x}|\theta^{(I_{n})})}{Z(\theta^{(I_{n})})}$.\\

~~~~ 2. Decide where to visit in the next iteration:
~~~~ \State $I_{n+1} \sim \lbrace 1,...,d \rbrace$ with probabilities $h(\mathbf{x}_{n+1}|\theta^{(i)})\exp(-c_{n}^{(i)})$ for $i=1,...,d$. \\

~~~~ 3. Update approximation of $\log{Z(\theta^{(i)})}$: Following stochastic approximation 

~~\citep{wang2001efficient,atchade2004wang}
~~~~ \State $c_{n+1}^{(i)} = c_{n}^{(i)}+\gamma_{n}\frac{h(\mathbf{x}_{n+1}|\theta^{(i)})\exp(-c_{n}^{(i)})}{\sum_{j=1}^{d} h(\mathbf{x}_{n+1}|\theta^{(j)})\exp(-c_{n}^{(j)})}$ for $i=1,...,d$, and $\gamma_{n+1}=\gamma_{n}$.

\EndWhile\\

Step size become smaller: $\gamma_{n+1} = \gamma_{n}/2$

\EndWhile\\

Return $\lbrace \mathbf{x}_{\tau},I_{\tau},\mathbf{c}_{\tau} \rbrace$, where $\tau$ is the total number of iterations for Part 1.
\end{algorithmic}
\end{algorithm}

\begin{algorithm}
\ContinuedFloat
\caption{ALR algorithm \textbf{(Part 2)}: Update the entire process. }\label{Atchadealgp2}

\begin{algorithmic}[H]

\normalsize

\State Initialize $\lbrace \mathbf{x}_{0},I_{0},\mathbf{c}_{0} \rbrace=\lbrace \mathbf{x}_{\tau},I_{\tau},\mathbf{c}_{\tau} \rbrace$ from Part 1 and set $H_{0}=\lbrace \mathbf{x}_{0},I_{0} \rbrace$.

\State For $(n+1)$st update, obtain $\lbrace \mathbf{x}_{n+1},I_{n+1},\mathbf{c}_{n+1} \rbrace$ using Step 1-3 in the Part 1 with deterministic step size, $\gamma_{n+1}=\epsilon_{1}/(n+1)^{0.7}$. 

\State Then append dataset: $H_{n+1}=H_{n} \cup \lbrace \mathbf{x}_{n+1},I_{n+1} \rbrace$.

\State 4. Approximate $Z(\theta)$ adaptively using all previous samples ($H_{n+1}$) and $\mathbf{c}_{n+1}$:

$\widehat{Z}_{n+1}(\theta)=\sum_{i=1}^{d}k(\theta,\theta^{(i)})\exp(c_{n+1}^{(i)})\frac{\sum_{k=1}^{n+1}\frac{h(\mathbf{x}_{k}|\theta)}{h(\mathbf{x}_{k}|\theta^{(i)})}1 \lbrace I_{k}=i \rbrace}{\sum_{k=1}^{n+1}1 \lbrace I_{k}=i \rbrace}$.

\State 5. Propose $\theta'\sim q(\cdot|\theta)$ and accept $\theta_{n+1}=\theta'$ with probability

$\alpha = \min\left\lbrace 1, \frac{p(\theta')h(\mathbf{x}|\theta')\widehat{Z}_{n+1}(\theta_{n})q(\theta_{n}|\theta')}{p(\theta_{n})h(\mathbf{x}|\theta_{n})\widehat{Z}_{n+1}(\theta')q(\theta'|\theta_{n})}\right\rbrace$, else reject (set $\lbrace \theta_{n+1},\mathbf{y}_{n+1}\rbrace=\lbrace \theta_n, \mathbf{y}_n \rbrace$).

\end{algorithmic}

\end{algorithm}

We split the ALR algorithms into two parts (see Figure~\ref{ALRFig}. The purpose of Part 1 of the Algorithm~\ref{Atchadealgp1} is to obtain a reasonable starting value for the normalizing function approximation without the memory and computational costs associated with updating the entire process $\lbrace \mathbf{x}_{n},I_{n},\mathbf{c}_{n},\theta_{n} \rbrace$. The step size $\lbrace \gamma_{n} \rbrace$ plays an important role in the convergence of $\mathbf{c}_{n}$ (analogous to how the gain factor $a_{n}$ is important for the abundance factor $\mathbf{w}_{n}$ in AEX). If $\gamma_{n}$ is too small, it takes a long time for convergence. If $\gamma_{n}$ is too large it can lead to large variance of $\mathbf{c}_{n}$. Therefore in the beginning of the algorithm, $\gamma_{n}$ is set to be a large value so that it can move around the space fast; it slowly becomes smaller after moving around the state space. In Algorithm~\ref{Atchadealgp1} (Part 1), the number of visits to each particle is recorded through $\mathbf{v}$. If $\mathbf{v}$ is close to uniform distribution ($\lbrace I_{n} \rbrace$ has visited $\lbrace 1,...,d \rbrace$ about equally), $\gamma_{n}$ becomes smaller. If step size $\gamma_{n}$ is smaller than some convergence criteria $\epsilon_{1}$, we can assume $\mathbf{c}_{n}$ has converged to a reasonable value.

The strengths and weakness of ALR are similar to those of AEX: ALR is asymptotically exact without requiring perfect sampling. However, the computational and memory costs can be large. Without low-dimensional sufficient statistics, the entire $\mathbf{x}_{n}$ chain needs to be stored with each iteration to calculate $\widehat{Z}_{n+1}(\theta)$ in Step 4 of the Algorithm~\ref{Atchadealgp2}. Furthermore, without sufficient statistics, calculations in Step 3 and Step 4 become expensive, because $h(\mathbf{x}_{k}|\theta),h(\mathbf{x}_{k}|\theta^{(i)})$ need to be recalculated; with sufficient statistics, one can simply take the product of $\theta$ or $\theta^{(i)}$ and the sufficient statistic of $\mathbf{x}_{k}$.

\textbf{Components to be tuned}: In this algorithm, step size ($\gamma_{n}$), the convergence checking components ($\epsilon_{1},\epsilon_{2}$), number of MH updates ($m$), number of particles ($d$) and kernel $k(\theta,\theta^{(i)})$ need to be tuned. \cite{atchade2008bayesian} set initial $\gamma_{0}$ as 1, $\epsilon_{1}=0.001,\epsilon_{2}=0.2$. Under these settings, consider the sequence of bounded stopping times $0=\tau_{0}<\tau_{1}<...<\tau_{10}=\tau$, where $\tau$ is the total number of iteration in Algorithm~\ref{Atchadealgp1} (Part 1). Until $\tau_{1}$, initial $\gamma_{0}$ is used as step size. For this step size, $\tau_{1}$ is the stopping time until each particle has been visited equally (i.e. $\max_{i=1,...,d}|v(i)-\bar{v}|\leq \epsilon_{2}\bar{v}$). Then $\gamma_{\tau_{1}+1}$ becomes $\gamma_{0}/2=1/2$ and is kept until $\tau_{2}$. This is repeated until $\tau_{10}$ where $\gamma_{\tau_{10}+1}=1/2^{10}<0.001=\epsilon_{1}$. Likewise, step size is controlled to hasten convergence of $\mathbf{c}_{n}$. Once $\mathbf{c}_{n}$ appears to have converged, Algorithm~\ref{Atchadealgp2} (Part 2) is implemented. $\gamma_{n+1}$ is updated as the deterministic sequence $0.001/(n+1)^{0.7}$ in Algorithm~\ref{Atchadealgp2} (Part 2). Similar to AEX, $m=1$ cycle of MH updates are enough in practice. For the number of particles, $d = 100p$ appears to work well in practice to cover the $p$-dimensional parameter space. Although various choice of kernel $k(\theta,\theta^{(i)})$ is possible, uniform kernel with bandwidth $h$ is used in this manuscript. This kernel gives $1/h$ weights for the $h$ closest particles and gives 0 weights for others. Bandwidth should be determined by trials and errors and we used $h = 20$. 

\textbf{Theoretical justification}: The ALR is an asymptotically exact algorithm and the ergodic average from the chain satisfies the Strong Law of Large Numbers. \cite{atchade2008bayesian} provide a proof that $\lbrace \theta_{n}\rbrace$ from the generated process converges to the target distribution exactly. Three assumptions are required: (1) $h(\mathbf{x}|\theta)$ is bounded away from 0 and $\infty$, (2) $q^{n_{0}}(\theta'|\theta) \geq \epsilon$ for all $\theta,\theta' \in \Theta$ where $q$ is proposal density and $n_{0} \geq 1 $ is an integer, (3) $\lbrace \gamma_{n} \rbrace$ is random and adaptively updated based on the previous generated process, which satisfies $\sum\gamma_{n}=\infty$ and $\sum\gamma^{2}_{n}<\infty$ almost surely. The first assumption holds for finite $\mathcal{X}$ and realistic parameter settings. The second assumption also holds for most symmetric kernels. The third assumption is technical, and taken from \cite{wang2001efficient}. Because we can typically control the sequence $\lbrace \gamma_{n}\rbrace$, this assumption is also generally easy to satisfy. Therefore, the assumptions appear to hold in most cases in practice.

\subsection{Pseudo-marginal MCMC}

~~~~~Suppose $\widehat{L}(\theta|\mathbf{x})$ is a positive and unbiased Monte Carlo approximation of $L(\theta|\mathbf{x})$. An MCMC algorithm that uses $\widehat{L}(\theta|\mathbf{x})$ in place of $L(\theta|\mathbf{x})$ is called pseudo-marginal MCMC \citep{beaumont2003estimation, andrieu2009pseudo} and its stationary distribution is equal to the desired target posterior distribution.


The pseudo-marginal framework in our context requires an unbiased approximation for $1/Z(\theta)$. An unbiased approximation of $Z(\theta)$ can be obtained using importance sampling estimate $\widehat{Z}(\theta)$ via MCMC samples from the likelihood. However, $1/\widehat{Z}(\theta)$ is a consistent but biased approximation for $1/Z(\theta)$. The Russian Roullette algorithm \citep{lyne2015russian} addresses this bias through a clever  geometric series correction (details are in the supplement). This is an asymptotically exact algorithm and the theoretical assumptions are generally satisfied for general forms of the probability model  $h(\mathbf{x}|\theta)$. However, it requires multiple $\widehat{Z}_{i}(\theta)$s at each iteration and each $\widehat{Z}_{i}(\theta)$ approximation itself requires multiple MCMC samples from $h(\mathbf{x}|\theta)$, making it computationally very expensive.

\subsection{Noisy MCMC and Hybrids}

~~~~~If a Markov chain with transition kernel $P$ satisfies detailed balance with respect to the target $\pi$, it is asymptotically exact. However, when $P$ is approximated by $\widehat{P}$, the samples generated may only approximately follow the target $\widehat{\pi}$. Such algorithms may be generically referred to as "noisy MCMC". 
 \cite{alquier2014noisy} describes a broad class of noisy MCMC algorithms and uses total variational distance to quantify the distance between the asymptotically exact and inexact chain. Noisy MCMC in its broader sence is a large class of algorithms, including, for instance, pseudo-marginal MCMC and ALR. Here we discuss a noisy MCMC algorithm that builds upon the exchange algorithm; hence this algorithm may be thought of as a hybrid between auxiliary and likelihood approximation-based methods.
 In the exchange algorithm, a single auxiliary variable $\mathbf{y}$ is generated from $h(\mathbf{y}|\theta')/Z(\theta')$. Therefore the $h(\mathbf{y}|\theta)/h(\mathbf{y}|\theta')$ term in the  \eqref{e5} may be thought of as a one-sample unbiased importance sampling estimate of $Z(\theta)/Z(\theta')$. Instead of using a single $\mathbf{y}$, if multiple auxiliary variables $\lbrace \mathbf{y}_{1},...,\mathbf{y}_{N} \rbrace$ are generated from the $h(\mathbf{y}|\theta')/Z(\theta')$ the resulting importance sampling approximation will have smaller variance. This is called the noisy exchange algorithm with acceptance probability

\begin{equation}
\alpha=\min\left\lbrace 1, \frac{p(\theta')h(\mathbf{x}|\theta')q(\theta|\theta')}{p(\theta)h(\mathbf{x}|\theta)q(\theta'|\theta)}\frac{1}{N}\sum_{i=1}^{N}\frac{h(\mathbf{y}_{i}|\theta)}{h(\mathbf{y}_{i}|\theta')}
\right\rbrace.
\label{e20}
\end{equation} 


 For $1< N < \infty$, the algorithm is asymptotically inexact because the detailed balance condition does not hold. \cite{liang2013monte} also proposes a similar approach called Monte Carlo MH (MCMH). \cite{alquier2014noisy} reports that the noisy exchange algorithm shows better mixing than the exchange algorithm, and proposes a noisy Metropolis adjusted Langevin (MALA) exchange algorithm, where the gradient of the intractable distribution is approximated to construct proposals, that further improves upon mixing. Also, the estimate for the ergodic average has smaller bias than the exchange algorithm in empirical studies. Of course, there may be a tradeoff between improving mixing and increasing computational costs per iteration. We skip details but mention the general noisy MCMC approach as it may lead to other useful algorithms. For instance, DMH with multiple auxiliary variables, say "noisy DMH", would be a simple extension. 

\textbf{Components to be tuned}: The number of auxiliary variables, $N$, is an additional component to be tuned compared to the exchange algorithm or DMH. The choice of $N$ depends on some tradeoffs: as $N$ becomes large, the constructed chain can have better mixing and estimates from the chain can have lower variance at the expense of computing time. Parallel computing may be helpful for sampling $N$  $\mathbf{y}_{i}$s, and evaluating $h(\mathbf{y}_{i}|\theta)/h(\mathbf{y}_{i}|\theta')$ independently.

\textbf{Theoretical justification}: \cite{alquier2014noisy} provides an upper bound for total variation norm distance between $P$, transition kernel for the exchange algorithm, and $\widehat{P}$, transition kernel for the noisy exchange algorithm, based on a result in  \cite{mitrophanov2005sensitivity}[Corollary 3.1], which requires uniform ergodicity of $P$. Two assumptions are used in this derivation: (1) the prior $p(\theta)$ is bounded away from $0$ and $\infty$, (2) $q(\theta'|\theta)$ is bounded away from $0$ and $\infty$; both assumptions are typically satisfied.

\section{Simulated and Real Data Examples}

~~~~~We now study the algorithms in the context of three models that are of general interest: (1) the Ising model, (2) a social network model, and (3) an attraction-repulsion point process model. These models also present different computational challenges. The code for this is implemented in {\tt R} \citep{ihaka1996r} and {\tt C++}, using the Rcpp and RcppArmadillo packages \citep{eddelbuettel2011rcpp}. We use the examples to compare the efficiency as well as practical implementation challenges of the algorithms. Simulation settings for the Russian roulette algorithm are provided in the supplementary material. 

\subsection{The Ising Model}

~~~~Consider an Ising model \citep{lenz1920beitrag,ising1925beitrag} on an $m$ by $n$ lattice  with parameter $\theta$. The observed lattice $\mathbf{x}=\{x_{ij}\}$ has binary values $x_{i,j}=\lbrace -1,1 \rbrace$, where $i,j$ denotes row and column location in the lattice. The model is

 \begin{equation}
L(\theta|\mathbf{x})=\frac{1}{Z(\theta)}\exp\left\lbrace \theta S(\mathbf{x}) \right\rbrace, ~~ S(\mathbf{x})=\sum_{i=1}^{m}\sum_{j=1}^{n-1}x_{i,j}x_{i,j+1} + \sum_{i=1}^{m-1}\sum_{j=1}^{n}x_{i,j}x_{i+1,j}, 
\label{e24}
\end{equation}

where spatial dependence is imposed via $S(\mathbf{x})$. The larger $\theta$ becomes, the stronger the interactions between the data on the lattice. Summation over all $2^{mn}$ possible configurations of this model is required for the calculation of the normalizing function, which is computationally expensive even for lattices of moderate size. The simulations are conducted using perfect sampling \citep{propp1996exact} on a 10$\times$10 lattice with uniform prior [0,1] with parameter settings representing moderate dependence, with $\theta=0.2$, and strong dependence, with $\theta=0.43$.

\begin{table}
\centering
\begin{tabular}{cccccccc}
 \hline
$\theta=0.2$ & Mean & 95\%HPD & ESS & Acc & Time(second) & ESS/Time \\
  \hline
AVM & 0.20 & (0.08,0.32) & 1527.88 & 0.37 & 12.46 & 122.62\\
  Exchange & 0.20 & (0.08,0.33) & 2061.23 & 0.49 & 15.29 & 134.81 \\
  DMH & 0.21 & (0.08,0.34) & 2068.29 & 0.48 & 1.78 & 1161.96 \\
  AEX & 0.21 & (0.07,0.35) & 1778.15 & 0.47 & 23.72 & 74.96\\
  ALR & 0.20 & (0.08,0.34) & 3647.34 & 0.59 & 9.33 & 390.93 \\
  RussianR & 0.20 & (0.06,0.33) & 2650.83 & 0.50 & 13849.10 & 0.19 \\
  NoisyDMH & 0.20 & (0.06,0.33) & 3347.76 & 0.59 & 71.02 & 47.14\\
  Gold & 0.20 & (0.08,0.33) & 9845.89 &  & & \\
   \hline
\end{tabular}
\begin{tabular}{cccccccc}
$\theta=0.43$ & Mean & 95\%HPD & ESS & Acc & Time(second) & ESS/Time\\
  \hline
AVM & 0.44 & (0.35,0.57) & 1020.12 & 0.32 & 10395.91 & 0.10\\
  Exchange & 0.43 & (0.33,0.54) & 2146.05 & 0.40 & 4889.32 & 0.44\\
  DMH & 0.47 & (0.34,0.61) & 2029.24 & 0.48 & 1.72 & 1179.79 \\
  AEX & 0.43 & (0.32,0.53) & 2048.01 & 0.41 & 24.96 & 82.05 \\
  ALR & 0.43 & (0.32,0.53) & 4125.59 & 0.51 & 9.88 & 417.57\\
  RussianR & 0.44 & (0.33,0.54) & 2708.55 & 0.45 & 30247.70 & 0.09 \\
  NoisyDMH & 0.47 & (0.32,0.61) & 3486.99 & 0.62 & 75.17 & 46.39 \\
  Gold & 0.43 & (0.33,0.54) & 10000.00 &  &  & \\
   \hline      
\end{tabular}
\caption{Inference results for an Ising model on a 10$\times$10 lattice. 20,000 MCMC samples are generated from each algorithm. The highest posterior density (HPD) is calculated by using coda package in {\tt R}. The calculation of Effective Sample Size (ESS) follows \cite{kass1998markov,robert2013monte}. "Acc" represents acceptance probability.}
\label{isingout}
\end{table}

All algorithms are tuned according to the methods described in the previous section. $h(\mathbf{y}|\widehat{\theta})/Z(\widehat{\theta})$ is chosen as the conditional density of the auxiliary variable $\textbf{y}$ for the AVM algorithm, where $\widehat{\theta}$ is MPLE. The auxiliary variable is generated by  10 cycles of Gibbs updates in DMH. For AEX, 100 particles are selected among 5,000 samples from fractional DMH (descriptions are in the supplementary material). Then the preliminary run of the auxiliary chain is implemented for 420,000 iterations with the first 20,000 iterations discarded for burn-in. To expedite computing and reduce memory management issues, the resulting samples are thinned (at equally spaced intervals of 20) to obtain 20,000 samples. In the auxiliary chain, we set $n_{0}=20,000$, $K_{s}=[0,100^{100+10s}]^{100}$, and $\mathbf{x}_{n}$ is updated by a single cycle of Gibbs updates. For ALR, 100 particles are drawn from the uniform prior and $\mathbf{x}_{n}$ is updated by a single cycle of Gibbs updates. The kernel giving equal weights for $h=10$ nearest particles and 0 for others is used. In noisy DMH, 100 samples are used to produce importance sampling estimate and same inner sampler is constructed as DMH. We note that in order to
make the computations feasible, we used parallel computing to obtain importance sampling estimates for both noisy DMH and the Russian Roulette algorithms. The parallel computing was implemented through OpenMP with the samples generated in parallel across 8 processors. We treated a run from the exchange algorithm as our gold standard; it was run for 1,010,000 iterations with 10,000 discarded for burn-in and 10,000 thinned samples are obtained from the remaining 1,000,000. Same simulation settings are used for $\theta=0.43$ case. Since perfect sampling takes very long for $\theta=0.43$ case, we use the ALR algorithm as the gold standard. All algorithms were run until the Monte Carlo standard errors  calculated by batch means \citep{jones2006fixed,flegal2008markov} are below 0.01.

\begin{figure}
\begin{center}
\includegraphics[scale=0.51]{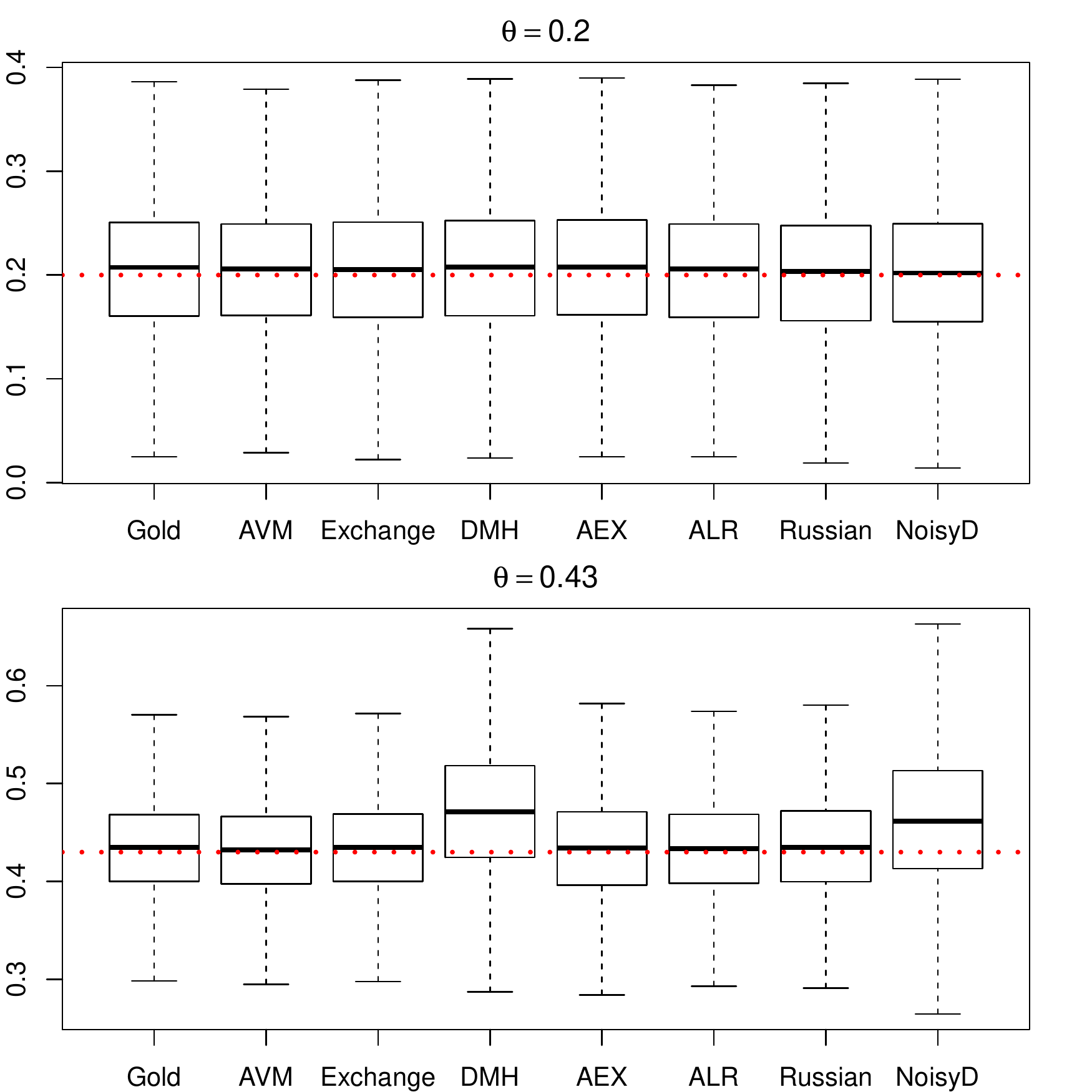}
\end{center}
\caption[]{Posterior densities for different $\theta$. Dotted line indicates true value.}
\label{isingplot}
\end{figure}

Table~\ref{isingout} shows that the estimates from different algorithms are well matched to the gold standard when $\theta=0.2$. Posterior densities in Figure~\ref{isingplot} also indicate agreement. Likelihood approximation approaches show larger ESS than auxiliary variable approaches. On the other hand, auxiliary variable approaches have smaller computational costs than likelihood approximation approaches. In particular, DMH shows the shortest computing time. For this parameter setting, inference results from both asymptotically exact and inexact algorithms are accurate.

However when $\theta=0.43$, inference results are biased for asymptotically inexact algorithms. In Table~\ref{isingout} and Figure~\ref{isingplot}, it is observed that the inference results from DMH and Noisy DMH do not match the gold standard. Due to the strong dependence at this parameter setting, mixing of the inner sampler to generate samples from the $h(\mathbf{x}|\theta)/Z(\theta)$ is slower than for the $\theta=0.2$ case. Therefore a large number of Gibbs updates are necessary for accurate inference. Although Noisy DMH provides a closer estimate to the true value than DMH because it uses multiple samples for estimating $Z(\theta)/Z(\theta')$, it is still biased. The number of iterations for the inner sampler in the Russian Roulette algorithm also has to be increased due to the slow mixing (see supplementary material for details). We can also observe that AVM and the exchange algorithm take several hours, because perfect sampling takes longer to achieve coalescence. On the other hand, there are no extra costs for AEX and ALR in the presence of strong dependence. For both algorithms, still a single cycle of Gibbs update is enough for updating $\mathbf{x}_{n}$. This is because both algorithms require updating not sampling $\mathbf{x}_{n}$ from the probability model.

\textbf{Summary}: In the case where the Ising model has only moderate dependence, all the algorithms work reasonably well. Here, DMH may be preferable because it is the easiest to construct and the fastest to run. On the other hand, in the context of Ising models that exhibit strong dependence, the computational problem becomes more challenging. For the asymptotically inexact algorithms like DMH, in order to obtain accurate results in the presence of strong dependence in the Ising model, the number of iterations for the inner sampler needs to be very large. This results in a much larger computational burden. AVM and the exchange algorithms also become impractical because perfect sampling takes too long. AEX and ALR can be useful for spatial autologistic models because both algorithms can attain accurate estimates with moderate computing speed even in the presence of strong dependence.  

\subsection{Social Network Models}

~~~~~Exponential random graph models (ERGM) \citep{robins2007introduction, hunter2008ergm} provide an approach for modeling relationships among nodes of a network. Consider the undirected ERGM with $n$ nodes. For all $i \neq j$, $x_{i,j}=1$ if the $i$th node and $j$th node are connected, otherwise $x_{i,j}=0$ and $x_{i,i}$ is defined as 0. This forms an $n$ by $n$ adjacency matrix, $\mathbf{x}$. Calculation of the normalizing function requires summation over all $2^{n(n-1)/2}$ configuration, which is infeasible. Here we investigate two ERGM examples, the first with 4 parameters and the second with 9 parameters; the latter example also involves a more complicated summary statistic which poses an additional computational challenge. We have not implemented AVM and the exchange algorithms because perfect samplers are possible only for a few special cases \citep{butts2012perfect}. 

\subsubsection{A Florentine Business Network}

~~~~In this example, we study the Florentine business dataset \citep{flobusiness}, which describes the business networks among 16 Florentine families. Consider the undirected ERGM, where the probability model is

\begin{equation}
L(\theta|\mathbf{x})=\frac{1}{Z(\theta)}\exp\left\lbrace \theta_{1}S_{1}(\mathbf{x}) + \theta_{2}S_{2}(\mathbf{x}) + \theta_{3}S_{3}(\mathbf{x}) + \theta_{4}S_{4}(\mathbf{x}) \right\rbrace,
\label{e25}
\end{equation}

\begin{gather*}
S_{l}(\mathbf{x})=\sum_{i=1}^{n}{x_{i+} \choose l},l=1,2,3; ~~~~ S_{4}(\mathbf{x})=\sum_{i<j<k}x_{i,j}x_{j,k}x_{k,i}.
\end{gather*}

\begin{table}[tt]
\centering
\begin{tabular}{cccccccc}
  \hline
$\theta_{2}$ & Mean & 95\%HPD & ESS & Acc & Time(second) & ESS/Time\\
  \hline
DMH & 1.24 & (0.02,2.57) & 1026.67 & 0.24 & 10.45 & 98.25 \\
AEX & 1.24 & (0.17,2.58) & 991.87 & 0.20 & 126.46 & 7.84 \\
ALR & 1.25 & (0.16,2.52) & 1456.57 & 0.33 & 2500.37  & 0.58 \\
RussianR &  1.27 & (0.03,2.68) & 1433.60 & 0.31 & 33534.96 & 0.04\\
NoisyDMH & 1.28 & (0.03,2.59) & 1600.90 & 0.32 & 297.35 & 5.38\\
Gold & 1.27 & (0.08,2.50) & 9655.90 &  &  & \\
   \hline
\end{tabular}
\caption{Inference results for 2-star in ERGM for Florentine business dataset. 30,000 MCMC samples are generated from each algorithm. The highest posterior density (HPD) is calculated by using coda package in {\tt R}. The calculation of Effective Sample Size (ESS) follows \cite{kass1998markov,robert2013monte}. "Acc" represents acceptance probability.}
\label{ergmout}
\end{table}

Sufficient statistics $S(\mathbf{x})=\left\lbrace S_{1}(\mathbf{x}),S_{2}(\mathbf{x}),S_{3}(\mathbf{x}),S_{4}(\mathbf{x}) \right\rbrace$ represent  the number of edges, two-stars, three-stars and triangles respectively, where $x_{i+}$ indicates row sum of adjacency matrix. Triangle represents the number of cyclic relationships and $k$-star indicates the number of nodes which have exactly $k$ relationships. 

Each of the algorithms is tuned according to the guidelines in the previous sections. The auxiliary variable is generated by 10 cycles of Gibbs updates in DMH. In AEX, 200 particles are selected among 5,000 samples from fractional DMH, as described in the supplementary material. Then the preliminary run of the auxiliary chain is implemented for 630,000 iterations with the first 30,000 iterations discarded for burn-in. The resulting samples are thinned (at equally spaced intervals of 20) to obtain 30,000 samples. In the auxiliary chain, we set $n_{0}=20,000$, $K_{s}=[0,100^{100+10s}]^{200}$, and $\mathbf{x}_{n}$ is updated by a single cycle of Gibbs updates. For ALR, 400 particles are chosen from the  short run of DMH, and a single cycle of Gibbs updates are used to update $\mathbf{x}_{n}$. The same kernel with $h=20$ is used as previous example. 100 samples are used for the importance sampling estimate in noisy DMH. Importance sampling estimates in  Russian Roulette and noisy DMH algorithms are obtained in parallel as in the previous example. To obtain a gold standard for comparisons, we run the AEX algorithm 10 times independently. Each run consists of 101,000 iterations with 1,000 iterations discarded as burn-in. Then 10,000 samples are obtained by thinning from the remaining 10 sets of 100,000 samples. All algorithms were run until the Monte Carlo standard error is at or below 0.02.

Here we only provide the inference results regarding $\theta_{2}$ because similar results are observed for the other parameters. Table~\ref{ergmout} indicates that the estimates from the different algorithms are similar to the those of the gold standard. This is because 10 cycles of Gibbs updates are enough to generate auxiliary samples from the likelihood in this example. However, for fewer iterations, say 1 to 2 cycles, the asymptotically inexact algorithms are biased. As in the Ising model example, less correlated samples can be generated from likelihood approximation approaches, at additional computational expense. Both ALR and AEX are computationally efficient compare to the Russian Roulette algorithm. This is because we can effectively store the previous samples using sufficient statistics in the likelihood. However, ALR is relatively expensive here (unlike in the Ising model), because it takes longer to visit the entire state space equally with increasing dimensions.  The performance of AEX is relatively robust in the multidimensional case and is the fastest among asymptotically exact algorithms if the particles are carefully chosen. 

\textbf{Summary}: AEX is the most reliable approach because it is asymptotically exact while at the same time retaining some computational efficiency because this model has low-dimensional statistics. On the other hand, DMH is simple and computationally efficient and as long as the length of the inner sampler is reasonable, it also provides accurate inference.

\subsubsection{An Emergent Multi-organizational
Network (EMON)}

~~~~In this example, we study the Mount St. Helens emergent multi-organizational network (EMON) dataset \citep{drabek1981managing}, which describes communication networks for the search and rescue operations among 27 organizations. There are three node attributes in the model: (1) sponsorship level (city, county, federal, private, and state), (2) command rank score (higher score indicates higher rank), (3) whether headquarters are sited locally to the impact area or not (L/NL). Consider the undirected ERGM, where the probability model is

\begin{equation}
L(\theta|\mathbf{x})=\frac{1}{Z(\theta)}\exp\left\lbrace \sum_{l=1}^{9} \theta_{l}S_{l}(\mathbf{x})\right\rbrace,
\label{emonmodel}
\end{equation}

\begin{gather*}
S_{1}(\mathbf{x})=\sum_{i=1}^{n}{x_{i+} \choose 1},~~S_{2}(\mathbf{x})=\sum_{i<j<k}x_{i,j}x_{j,k}x_{k,i}\\
S_{3}(\mathbf{x})=\sum_{i<j}x_{i,j}(1\lbrace  \mbox{sponsor}_i=\mbox{county} \rbrace+1\lbrace \mbox{sponsor}_j=\mbox{county} \rbrace)\\
S_{4}(\mathbf{x})=\sum_{i<j}x_{i,j}(1\lbrace \mbox{sponsor}_i=\mbox{federal} \rbrace+1\lbrace \mbox{sponsor}_j=\mbox{federal} \rbrace)\\
S_{5}(\mathbf{x})=\sum_{i<j}x_{i,j}(1\lbrace \mbox{sponsor}_i=\mbox{private} \rbrace+1\lbrace \mbox{sponsor}_j=\mbox{private} \rbrace)\\
S_{6}(\mathbf{x})=\sum_{i<j}x_{i,j}(1\lbrace \mbox{sponsor}_i=\mbox{state} \rbrace+1\lbrace \mbox{sponsor}_j=\mbox{state} \rbrace)\\
S_{7}(\mathbf{x})=\sum_{i<j}x_{i,j}(\mbox{command}_i+\mbox{command}_j)\\
S_{8}(\mathbf{x})=\sum_{x_{i,j} \in L}GD_{4},~~S_{9}(\mathbf{x})=\sum_{x_{i,j} \in NL}GD_{4}
\end{gather*}

\begin{table}[tt]
\centering
\begin{tabular}{cccccccc}
  \hline
$\theta_{9} \times 10^2$ & Mean & 95\%HPD & ESS & Acc & Time(minute) & ESS/Time\\
  \hline
DMH & 1.93 & (0.61,2.98) & 583.08 & 0.05 & 16.10  &  36.21\\
NoisyDMH & 1.93 & (0.08,3.45) & 796.01  & 0.08 & 1026.15 & 0.78\\
Gold & 1.89  & (0.58,2.98) & 5494.31 &  &  & \\
   \hline
\end{tabular}
\caption{Inference results for graphlet orbit factor for "NL" in EMON dataset. 40,000 MCMC samples are generated from each algorithm. The highest posterior density (HPD) is calculated by using the coda package in {\tt R}. The calculation of Effective Sample Size (ESS) follows \cite{kass1998markov,robert2013monte}. "Acc"  represents acceptance probability.}
\label{ergmout2}
\end{table}

The sufficient statistics are $S_{1}(\mathbf{x})$ (the number of edges), $S_{2}(\mathbf{x})$ (triangles), $S_{3}(\mathbf{x}) - S_{6}(\mathbf{x})$ (node factor for sponsorship level), $S_{7}(\mathbf{x})$ (node covariance for command rank score), and $S_{8}(\mathbf{x})-S_{9}(\mathbf{x})$ (graphlet orbit factor for location of headquarters). Graphlet statistics are small, connected subgraphs which represent the certain topological structure of a network \citep{prvzulj2004modeling,prvzulj2007biological}. Here we used Graphlet 4 with automorphism orbit 7 for describing brokerage roles between the 27 organizations \citep{yaveroglu2014ergm}. 

In this model there are practical implementation issues for both particle-based algorithms (ALR and AEX): (1) Both algorithms require increasing number of particles to cover the high-dimensional parameter space. (2) The algorithms used for generating particles can exhibit slow mixing. For example the default implementation of DMH mixes slowly, which implies that it takes longer chains to generate the necessary particles for the ALR or AEX algorithms. (3) Even after the particles are generated, the stochastic approximation algorithm can still be slow when the number of parameters is large. Even with large gain factor (AEX) or step size (ALR), visiting each state space equally is infeasible when the parameter dimension is high (as in the 9-dimensional case above). This issue is perhaps even more problematic than issues (1) and (2) above.

The Russian Roulette algorithm is also computationally expensive because of the  complexity of summary statistics in this example (wall time of roughly 1 month). Therefore, we study only the DMH and noisy DMH algorithms for this example. The details of our implementation are as follows. We generate auxiliary variables via 10 cycles of Gibbs updates, and 300 samples are used to construct an importance sampling estimate in noisy DMH. Importance sampling approximations in noisy DMH are evaluated through parallel methods, as in previous examples. We use the DMH algorithm with 20 cycles as the gold standard; it was run for 101,000 iterations with 1,000 samples discarded for burn-in and 10,000 thinned samples are obtained from the remaining samples. All algorithms are run until the Monte Carlo standard error is no larger than 0.03.
 
Here we provide inference results for $\theta_{9}$ because similar results are observed for the other parameters. Table~\ref{ergmout2} shows that posterior means from the different algorithms are well matched to the gold standard. It is observed that the acceptance rate is much lower than  in the simpler Florentine business data example. The highest posterior density (HPD) interval obtained from Noisy DMH is slightly wider than that of the gold standard. As in the previous examples, a Noisy DMH algorithm can generate less correlated samples with higher acceptance rates compared to the DMH algorithm.

\textbf{Summary}: Particle-based algorithms (AEX and ALR) are infeasible for the 9-dimensional parameter case even though there are summary statistics to be stored. Because of computationally expensive graphlet statistics, Russian Roulette is infeasible. With the choice of an appropriate length for the inner sampler, DMH can provide accurate inference, with the highest ESS/T. This fact suggests that for these challenging cases, DMH may still be a practical approach.

\subsection{Spatial Interaction Point Process}

~~~~A spatial point process $\mathbf{x}=\lbrace x_{1},...,x_{n} \rbrace$ is a realisation of random points in a bounded plane $S \subset R^{2}$. By introducing an interaction function $\phi(D_{ij})$ which is the function of distance between the coordinates of $x_{i}$ and $x_{j}$, a probability model may be used to describe spatial patterns among the point. Extending the Strauss process \citep{strauss1975model} which explains repulsion patterns among point, \cite{goldstein2014attraction} develop a point process model to explain both attraction and repulsion patterns of the cells infected with human respiratory syncytial virus (RSV). The interaction function is

\begin{equation}
\phi(D) = \begin{cases}
      0 & 0 \leq D \leq R \\
      \theta_{1}-\left(\frac{\sqrt{\theta_{1}}}{\theta_{2}-R}(D-\theta_{2}) \right)^{2}  & R< D \leq D_{1} \\
      1+\frac{1}{(\theta_{3}(D-D_{2}))^{2}} & D > D_{1}
\end{cases}
\label{e26}
\end{equation}

and the probability model is 

\begin{equation}
L(\theta|\mathbf{x})=\frac{\lambda^{n} \left[\prod_{i=1}^{n} \exp\left\lbrace \min\left(\sum_{i\neq j}\log{(\phi(D_{i,j}))},1.2\right)  \right\rbrace \right]}{Z(\theta)}, ~~ \theta=\lbrace \lambda,\theta_{1},\theta_{2},\theta_{3} \rbrace.
\label{e27}
\end{equation}

There are four parameters $\lbrace \lambda,\theta_{1},\theta_{2},\theta_{3} \rbrace$ in the model: $\lambda$ controls intensity of the point process and $\lbrace \theta_{1},\theta_{2},\theta_{3} \rbrace$ are the parameters controlling the interaction function. $\theta_{1}$ is the peak value of $\phi$, $\theta_{2}$ is value of $D$ at the peak of $\phi$ and $\theta_{3}$ represents descent rate after the peak. When the points are too close to each other, $\phi$ value in \eqref{e26} is less than $1$ which means points have a tendency to remain apart. However as the distance between points is larger, $\phi$ value become increased which means that points clump together. Attraction patterns become smaller as the distance between the points become larger. Likewise, this model can capture attraction repulsion spatial association among infected cells. Calculation of the normalizing function requires integration over the continuous domain $S$, which is intractable.

\cite{goldstein2014attraction} implement DMH for three independent replicates of 3,200 points in a well with radius 1,350 pixels, which is about 10,000 points. Even with code written in {\tt C}, inference took roughly 12 hours. This is because the number of points $n$ is large, and for each iteration of the DMH, thousands of birth-death MCMC steps are required to generate the auxiliary variable. To allow for a comparison with other algorithms, which are computationally more expensive than DMH, we work with a smaller point pattern; however, this pattern is still computationally challenging enough to serve as a good testbed for the various algorithms we consider. Simulations are conducted on a well with 337.5 radius pixels and number of points $n\approx 200$ without replicates. Point process $\mathbf{x}$ is generated through a long run of birth-death MCMC. We follow RSV-B simulation settings in \cite{goldstein2014attraction}, where the true parameter is $\lbrace \lambda\times 10^4,\theta_{1},\theta_{2},\theta_{3} \rbrace=\lbrace 4,1.2,15,0.3 \rbrace$. We use uniform priors on $\theta$. Since the number of data points is small, descent rate parameter $\theta_{3}$ is not recovered well. Therefore $\theta_{3}$ is fixed at $0.3$ and we infer the other three parameters. 

\begin{table}[tt]

\centering

\begin{tabular}{cccccccc}

  \hline

$\theta_{1}$ & Mean & 95\%HPD & ESS & Acc & Time(minute) & ESS/Time\\

  \hline

DMH & 1.20 & (1.03,1.33) & 1343.11 & 0.27 & 3.93 & 341.35\\

  RussianR & 1.19 &(1.04,1.35) & 1867.54 & 0.32 & 3299.76 &  0.57\\

  NoisyDMH & 1.20 & (1.04,1.34) & 2676.49 & 0.42 & 77.73 & 34.44 \\

  Gold & 1.19 & (1.03,1.34) & 4397.39 &  &  & \\

   \hline

\end{tabular}
\caption{Inference results for outputs for $\theta_{1}$ in attraction repulsion point process model ($n \approx 200)$. 40,000 MCMC samples are generated from each algorithm. The highest posterior density (HPD) is calculated by using coda package in {\tt R}. The calculation of Effective Sample Size (ESS) follows \cite{kass1998markov,robert2013monte}. "Acc" represents acceptance probability.}

\label{attractionout} 

\end{table}

Implementing AVM or the exchange algorithm is infeasible because perfect sampling is impossible for this example. Although both AEX and ALR can be implemented theoretically, it is not practical because there are no summary statistics. Without summary statistics, we need to store the distance matrix of cumulative point process samples with varying dimension around 200 by 200, which is a burden on memory. Therefore, we study DMH, noisy DMH and the Russian Roulette algorithms for this example. Each algorithm is tuned according to the previous sections. For all the algorithms, samples are generated from the likelihood through 2,000 iterations of birth-death MCMC. For noisy DMH, 100 samples are used to construct an importance sampling approximation with each iteration. Importance sampling approximations in noisy DMH and Russian Roulette are evaluated through parallel methods, as in the past. We use the Russian Roulette algorithm as the gold standard; it was run for 101,000 iterations with 1,000 samples discarded for burn-in and 10,000 thinned samples are obtained from the remaining samples. All algorithms were run until the Monte Carlo standard error is at or below 0.01.

Here we only provide results for $\theta_{1}$ because similar results are observed for the other parameters. Table~\ref{attractionout} shows that inference results from the different algorithms are well matched to the gold standard and the highest posterior density (HPD) intervals cover true values. As in the previous examples, likelihood approximation approaches can generate less correlated samples with higher acceptance rates. 

\textbf{Summary}: DMH has significant advantages in terms of computational efficiency. Compared to the Russian Roulette algorithm which takes about 55 hours, DMH only takes several minutes to be implemented. This is because both calculation of the $h(\mathbf{x}|\theta)$ and the inner sampler to generate  auxiliary variables are expensive compared to previous examples. Furthermore, considering ESS/T, DMH can generate effective sample within the shortest time, while simulated samples are close to the gold standard. This fact demonstrates that for computationally expensive problem, especially the model without low-dimensional sufficient statistics, we recommend DMH as a practical approach, though asymptotically exact inference is not guaranteed theoretically. Considering that the example is 16 times smaller than the original, it is infeasible to implement other algorithms for the original problem.

\section{Computational Complexity}

~~~~~Here we investigate the computational complexity of the algorithms, that is, how the algorithms scale as we increase the number of data points. We believe this is particularly relevant in modern statistics since a common and important question to ask is how well an algorithm works as data sizes get large. Algorithms requiring perfect sampling will not be considered, because perfect sampling can only be constructed for limited cases. Let $L(\cdot)$ be the complexity of evaluating $h(\mathbf{x}|\theta)$, $G(\cdot)$ be the complexity of inner sampler, and $n$ be the number of points in data $\mathbf{x}$.

We begin with a few caveats. It is challenging to calculate complexity because some  algorithms contain random quantities (stopping times), or have many components that need to be tuned specifically for different problems. Also, the mixing of the Markov chains and the quality of asymptotically inexact algorithms are influenced by the data itself (for instance, dependence) and not just the size of the data. We also note that to simplify calculations, we assume the dimensions of the data $\mathbf{x}$ and the auxiliary variable $\mathbf{y}$ are assumed to be same ($n$). This is not strictly accurate for the point process example where the auxiliary variable is generated through birth-death MCMC with varying dimensions. However, the dimensions are approximately similar to original data's dimension $n$. In the point process example we can store data $\mathbf{x}$ itself or distance matrix of $\mathbf{x}$ in the adaptive algorithms (AEX, ALR) to evaluate $h(\mathbf{x}|\theta)$. The latter can avoid re-computations at the expense of memory. Here we assume that distance matrix of $\mathbf{x}$ will be stored. We assume that the length of the inner sampler is proportional to $n$. It is true when the inner sampler is used for update (AEX, ALR) which requires a single cycle of update (proportional to $n$). However, it may not be correct if the purpose of the inner sampler is sampling (DMH, Russian Roulette, Noisy DMH) from the probability model. The mixing of the inner sampler can be different depending on the parameter settings. However, several cycles of update (proportional to $n$) appears to work in practice.

\begin{figure}

\includegraphics[ scale = 0.6]{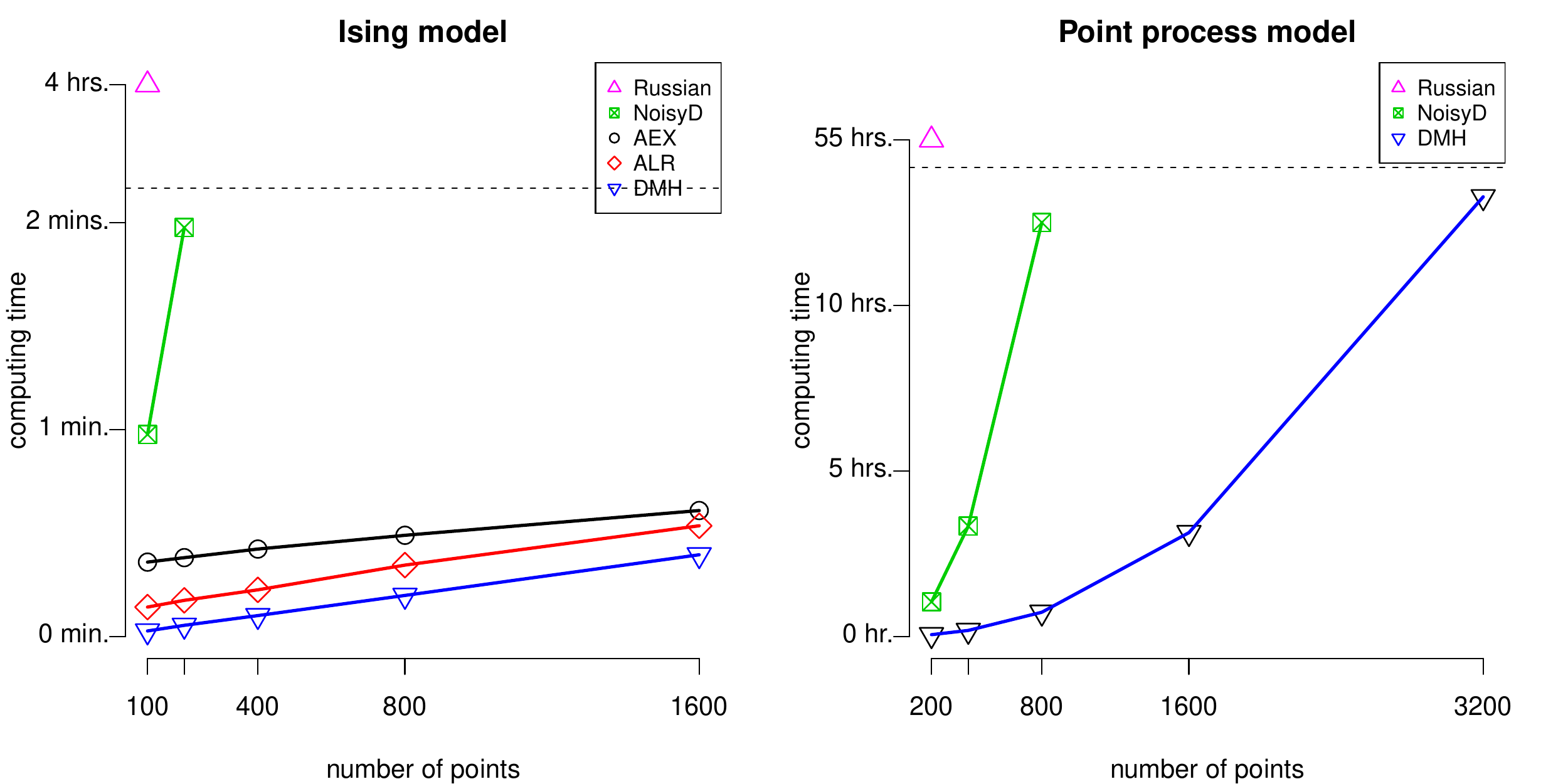}

\caption[]{Illustration of the observed computing time for algorithms. For Ising model, time is measured for $\theta=0.2$ with 20,000 iterations. For point process model, time is measured for RSV-B simulation settings in \cite{goldstein2014attraction} with 40,000 iterations.}

\label{complexityfigure}

\end{figure}

All the algorithms require sampling (DMH, Noisy DMH, Russian Roultte) or updating (AEX, ALR) from the probability model and evaluating $h(\mathbf{x}|\theta)$. In exponential family models such as Ising model or ERGM, complexity of $L(\cdot)$ is of order $n$, because the number of calculations for $S(\mathbf{x})$ is proportional to $n$. For the inner sampler, when a single point is proposed to be changed, only neighboring points (fixed number regardless of $n$) are affected due to the Markovian assumption. Because the length of the inner sampler is assumed to increase with $n$, $G(\cdot)$ is also of order $n$. On the other hand, complexity of $L(\cdot)$ for point process model is of order $n^2$, because evaluation of the $h(\mathbf{x}|\theta)$ requires calculating an $n$ by $n$ distance matrix for $n$ data points, and evaluating interaction function $\phi(\cdot)$ on the corresponding distance matrix. For the inner sampler, when a single point is proposed to be added (birth) or deleted (death), the distance of a proposed point from other $n$ points should be calculated, and then $\phi(\cdot)$ should be evaluated at each point. Since the length of the inner sampler is assumed to be proportional to $n$, $G(\cdot)$ is of order $n^2$.

There is a big difference in calculating $h(\mathbf{x}|\theta)$ for exponential family and point process model. For the exponential family model, once we evaluate $S(\mathbf{x})$, we can simply take the product of $\theta$ and $S(\mathbf{x})$ for evaluation of $h(\mathbf{x}|\theta)$ in different $\theta$. On the other hand for point process model, $h(\mathbf{x}|\theta)$ should be recalculated; $\phi(\cdot)$ should be evaluated at the distance matrix of $\mathbf{x}$ with different parameters. Here we provide our main observations (see supplement for details): (1) Complexity of exponential family is $\mathcal{O}(n)$ and $\mathcal{O}(n^2)$ in point process for all the algorithms. (2) Although algorithms have the same complexity, there are major differences in calculations per iteration. The amount of calculations per iteration of AEX, ALR and DMH are similar for exponential family models. Considering complexity and memory costs, DMH is the cheapest algorithm for both models. (3) Adaptive algorithms (AEX and ALR) require more memory. With increasing iterations, both algorithms become slower because algorithms use cumulative samples $H_{n+1}$ in calculations per iteration (AEX: Step 5 of the Algorithm~\ref{AEXalg2}, ALR: Step 4 of the Algorithm~\ref{Atchadealgp2}). Memory may not be an issue for the exponential family model. However for models without low-dimensional summary statistics, memory costs can become prohibitively expensive. 

Figure~\ref{complexityfigure} is the observed computing time for several algorithms with different scales in both models. We only include parts of results for Noisy DMH and the Russian Roulette algorithm. This is because both algorithms are too expensive to compare with other algorithms for large scales. For different scales, mixing of each algorithm is determined to be similar based on effective sample sizes, once we use appropriate step size (covariance) for proposal. Step size can be tuned to achieve similar acceptance rate for different scales; the larger data size becomes, the smaller step is used. We can also adaptively update step size \citep{atchade2006adaptive,atchade2008bayesian}. Figure~\ref{complexityfigure} supports our calculations about $\mathcal{O}(n)$ complexity for exponential family and $\mathcal{O}(n^2)$ complexity for point process model. Also in the exponential family model, slopes of AEX, ALR and DMH are similar to each other. However DMH is the fastest because of memory requirements of both adaptive algorithms. These facts are consistent with our calculations.

\section{Connections and Summary of Results}

~~~~~Here, we point out connections between the algorithms. Also, based on our study, we provide some conclusions about advantages and disadvantages of each algorithm.

\subsection{Connections and Observations}

\begin{figure}
\begin{center}
\includegraphics[scale=1.2]{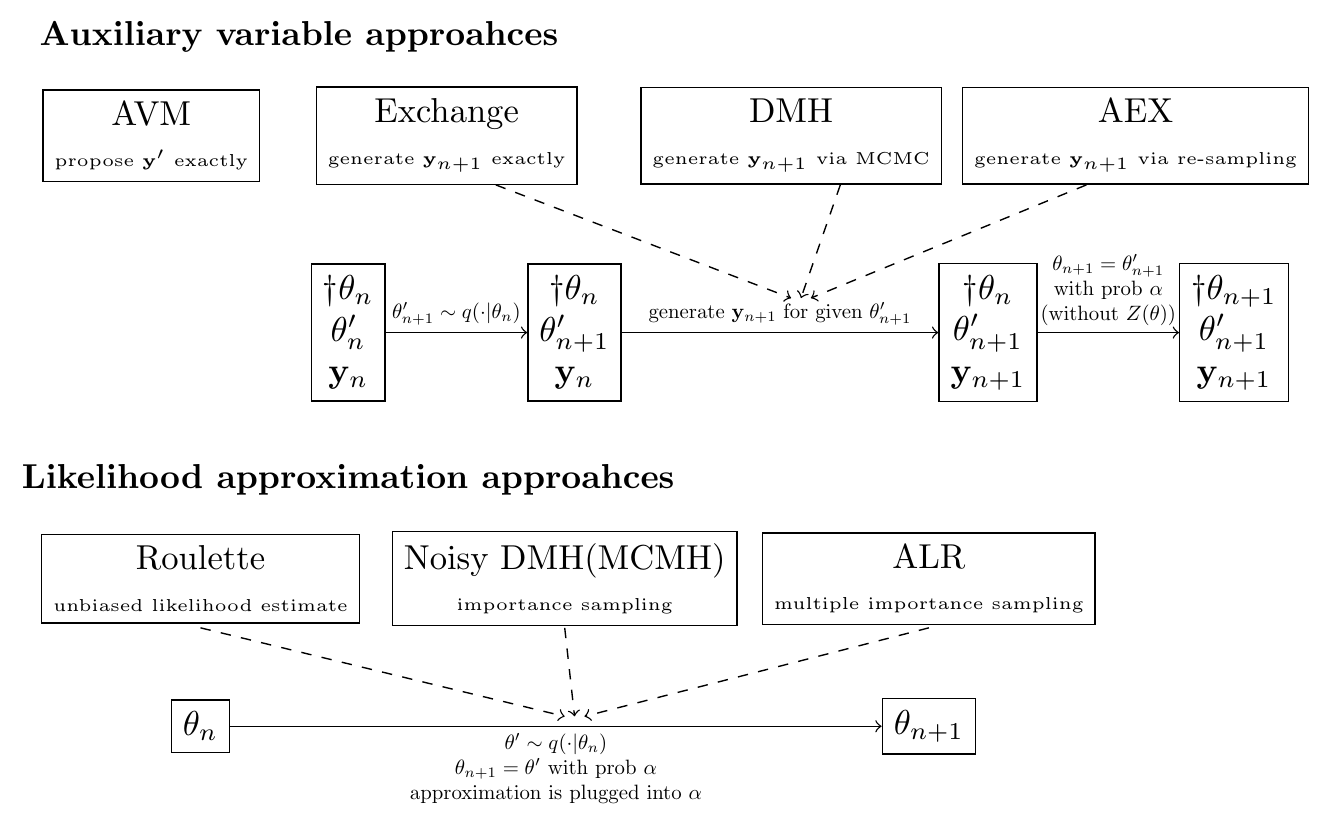}
\end{center}
\caption[]{Connections between the algorithms. The dagger symbols indicate the parameter of interest in auxiliary variable approaches.}
\label{ConnectionFig}
\end{figure}

~~~~~All the algorithms require sampling from the probability model either approximately (MCMC or particle methods) or exactly (perfect sampling). These samples, in turn, are used in some form of an importance sampling estimate. This is clear in the likelihood approximation approaches; for the auxiliary variable approaches one may think of the acceptance probability of the Metropolis-Hastings algorithm as containing a single sample importance sampling approximation of $Z(\theta)/Z(\theta')$. Let the conditional density of the auxiliary variable in AVM  be $f(\mathbf{y}|\theta,\mathbf{x})=h(\mathbf{y}|\widehat{\theta})/Z(\widehat{\theta})$, where $\widehat{\theta}$ is the MPLE. The part of the acceptance probability only related to the auxiliary variable is

\begin{equation}
\frac{f(\mathbf{y}'|\theta',\mathbf{x})h(\mathbf{y}|\theta)}{h(\mathbf{y}'|\theta')f(\mathbf{y}|\theta,\mathbf{x})}=\frac{h(\mathbf{y}'|\widehat{\theta})h(\mathbf{y}|\theta)}{h(\mathbf{y}'|\theta')h(\mathbf{y}|\widehat{\theta})}.
\label{e23}
\end{equation}

\cite{murray2006} point out that since $\mathbf{y}\sim h(\cdot|\theta)/Z(\theta)$ and $\mathbf{y}'\sim h(\cdot|\theta')/Z(\theta')$, $h(\mathbf{y}'|\widehat{\theta})/h(\mathbf{y}'|\theta')$ and $h(\mathbf{y}|\widehat{\theta})/h(\mathbf{y}|\theta)$ may be thought of as one-sample unbiased importance sampling approximations of $Z(\widehat{\theta})/Z(\theta')$ and $Z(\widehat{\theta})/Z(\theta)$ respectively. Therefore, \eqref{e23} is the ratio of two unbiased estimates which is a biased estimate of $Z(\theta)/Z(\theta')$. \cite{murray2006} explain that compared to AVM, the exchange algorithm is more direct because $h(\mathbf{y}|\theta)/h(\mathbf{y}|\theta')$ in the \eqref{e5} is one-sample unbiased importance sampling estimate for $Z(\theta)/Z(\theta')$ where $\mathbf{y}\sim h(\cdot|\theta')/Z(\theta')$. Though sampling schemes for the auxiliary variable are different, the same logic is applied to DMH and AEX. The only difference is that DMH generates $\mathbf{y}$ via MCMC, and AEX generates $\mathbf{y}$ via resampling (dynamic importance sampling from the mixture distribution). Both classes of algorithms are clearly connected through importance sampling. We summarize these connections in Figure~\ref{ConnectionFig}. 

 AEX lies at the intersection of both classes of algorithms. As in the likelihood approximation approach, in the auxiliary chain $w^{(i)}_{n}$ approximates (up to a constant) $Z(\theta^{(i)})$ for each $\theta^{(i)}$. As in the auxiliary variable methods, the intractable functions are cancelled in the target chain at each iteration. Furthermore, AEX is closely connected to ALR. Both algorithms approximate $Z(\theta^{(i)})$ at each particle -- $w_{n}^{(i)}$ in AEX and $c_{n}^{(i)}$ in ALR. Using their respective approximations, both collect a sample from a family of distributions $\lbrace h(\mathbf{x}|\theta^{(1)})/Z(\theta^{(1)}),...,h(\mathbf{x}|\theta^{(d)})/Z(\theta^{(d)}) \rbrace$ via (different) stochastic approximations and save a sample with each iteration. As the number of  iterations grows, the accumulated dataset $H_{n+1}$ grows, which makes both algorithm  asymptotically exact. In AEX the resampling distribution of $\mathbf{y}$ becomes close to $h(\mathbf{y}|\theta')/Z(\theta')$ and in ALR the approximation $\widehat{Z}_{n+1}(\theta)$ converges to the truth. The difference is that AEX cancels out $Z(\theta)$ while ALR plugs-in $\widehat{Z}_{n+1}(\theta)$ into the acceptance probability.

In general, likelihood approximation approaches have better mixing than auxiliary variable approaches. However auxiliary variable approaches are less expensive per iteration resulting in higher effective sample size per second. Hence, auxiliary variable approaches tend to often be faster than likelihood approximation approaches. However the efficiency of algorithms can change depending on the model, parameter settings, and the data, as well as decisions about tuning components in each algorithm. In particular, strong dependence in observations tends to,  slow sample generation, both with approximately and exact samplers. Perfect sampling takes longer to achieve coalescence, which can slow the AVM and exchange algorithm. Also, MCMC sampling requires longer chains, increasing  computing time for the DMH and the Russian Roulette algorithms as well. In this case, AEX and ALR show relatively robust performance because both algorithms require updating $\mathbf{x}_{n}$, not sampling $\mathbf{x}_{n}$ from the probability model. For both algorithms, the choice of particles is crucial in determining both statistical and computational efficiency. If the parameter has higher dimensions, fewer particles are likely to be in the higher density regions of the posterior distribution of $\theta$. Then the approximation of $Z(\theta)$ might be inaccurate (ALR) or resampled auxiliary variable might be improbable (AEX), which can lead to poor mixing. Requiring an increase in the number of particles also slows computing. Therefore, careful choice of particles is important (see the supplementary material), especially for multidimensional parameters.

\subsection{Guidelines and Recommendations}


~~~~Comparing asymptotically exact and inexact algorithms is challenging in general because we have to simultaneously account for two different aspects of the algorithms: (1) mixing of the Markov chain, which affects the rate at which sample-based approximations converge to the true values, and (2) the quality of the target approximation in the case of asymptotically inexact algorithms. In order to provide reasonable comparisons, we always check the final results in each case using a variety of diagnostics, for example plots of marginal distributions as the Monte Carlo sample size increases, in order to convince ourselves that the approximation we obtain finally is close enough to the truth so that the above issues are avoided. Once this is the case, we can compare the efficiency of the algorithms via effective sample size (ESS) and effective samples per unit time (ESS/T). We have used ESS as a practical criteria of comparing mixing of the different algorithms; ESS/T can measure the computational efficiency as well.


\begin{table}[hh]
\centering
\begin{tabular}{ccccccc}
\hline

        &        & Exact or  &            &               &  \\

Method  & Class  & inexact   & Adaptation & Requirement   & Ease of use \\

\hline

AVM & A & Exact & No & Perfect sampling & 3\\

Exchange & A &  Exact & No & Perfect sampling & 2\\

DMH &    A & Inexact  & No & Nothing  & 1\\

AEX & A/L & Exact & Yes & Sufficient statistics & 4 \\

ALR &    L & Exact & Yes & Sufficient statistics & 4\\

Russian Roulette & L & Exact & No & Nothing & 4\\

Noisy DMH &  L & Inexact & No & Nothing & 2\\

\hline
\end{tabular}
\caption{Comparison between algorithms}
\label{summary}
\end{table}

~~~~Table~\ref{summary} summarizes algorithms based on the following criteria:

\begin{itemize}
\item Class: which class the each algorithm falls in. A,L indicates auxiliary variable approaches and likelihood approximation approaches respectively and A/L indicates both. 

\item Exact or inexact: whether the stationary distribution of the Markov chain is exactly equal to the target posterior in the long run or not.

\item Adaptation: whether the algorithm is adaptive (stationary distribution changes) or not.

\item Requirement: key requirement(s) for the algorithm to be practical.

\item Ease of use (rank): how much users need to make decision to run the algorithm. The smallest number indicates the simplest one.

\end{itemize}

AVM and the exchange algorithm propose the novel idea of cancelling out $Z(\theta)$ and thus have inspired many other algorithms. However, it is difficult to implement both algorithms in practice because the algorithms require perfect sampling. Even if we can implement perfect sampling it still has limited applicability because perfect sampling is very slow, either in the presence of strong dependence or for large scale data.

AEX, ALR, and Russian Roulette are asymptotically exact without perfect sampling. AEX is the fastest, especially when there are low-dimensional sufficient statistics for the model. AEX is faster than ALR for multidimensional $\theta$ cases, because AEX doesn't have random stopping time ($\tau$). For ALR it is observed that $\tau$ in Algorithm~\ref{Atchadealgp1} can be large for multidimensional $\theta$ problems, because it becomes difficult to move around the state space equally. This makes ALR slower than AEX for the complex parameter space. Therefore we recommend AEX for the model with low-dimensional summary statistics such as spatial autologistic model and ERGM. However as we describe in Section 4.2.2, both algorithms become less practical as the parameter space increases.

Without low-dimensional summary statistics, AEX poses memory burdens. In principle, the Russian Roulette algorithm can be applicable for the general form of the likelihood without sufficient statistics. The Russian Roulette algorithm does not require summary statistics or the  assumption that $h(\mathbf{x}|\theta)$ should be bounded. However, it may not be practical because constructing unbiased likelihood estimate requires $\tau$ number of unbiased estimates $\widehat{Z}_{i}(\theta)$ per iteration. If $\widehat{Z}_{i}(\theta)$ requires $N$ samples, $\tau N$ samples need to be generated from the likelihood using inner sampler per iteration. For this reason, the Russian Roulette algorithm is computationally expensive compared to other methods.

We recommend DMH for computationally expensive likelihoods without low-dimensional summary statistics. Since only a single sample from the likelihood is required per iteration, DMH is computationally cheaper than any other algorithms. Furthermore, the algorithm is widely applicable without requiring any assumptions and relatively robust for high-dimensional parameter space. Once, we choose the appropriate number of the inner sampler updates, we can obtain plausible inference results in practice as well as the effective samples within the shortest time (ESS/T). However exactness is not guaranteed in DMH. Finally, if we can afford some additional computing expenses, noisy DMH may be useful because it can result in fast mixing chains. 


\section{Discussion}

~~~~~We have discussed MCMC algorithms for likelihoods with intractable normalizing functions, explaining some theoretical underpinnings as well as examining practical implementation issues and efficiencies. We find that auxiliary variable approaches tend to be more efficient than likelihood approximation approaches, though efficiencies vary quite a bit depending on the dimension of the problem, and on the availability of low-dimensional sufficient statistics for models (especially for AEX and ALR algorithms). The specifics of the dataset also matter, for instance data sets with strong dependence can slow down algorithms for inferring parameters of spatial point processes. There is some overlap among the central ideas for most algorithms and all have a clear connection to importance sampling.


For users, practical implementation challenges are as important as theoretical justifications. \cite{chopin2015leave} also point out ease of use (fewer components to be tuned) and generalizability of code as important criteria for comparing Bayesian computation algorithms. From this perspective, AEX and ALR may be difficult choices for practitioners though both algorithms are asymptotically exact and moderately fast. This is because users have to manually tune lots of components, and do so differently for different problems. AEX seems to be a good choice for social network examples, at least in exponential random graph models where models have low-dimensional sufficient statistics. However, even here, \cite{jin2013bayesian} discuss  the need to carefully tune AEX in order to address some well known degeneracy problems \citep{handcock2002statistical,schweinberger2011instability}.


Overall, we suggest starting with the DMH algorithm for its ease of implementation and the fact that it is typically computationally efficient. If instabilities in results are observed, for instance when the inner sampler length is increased, then AEX may be a more robust approach, even though it is much harder to implement. A preliminary run of DMH can be helpful for both tuning (e.g. selecting particles) and debugging of the AEX algorithm. Sophisticated users of MCMC and particle-based algorithms could also consider ALR as an alternative, and noisy MCMC methods may also be useful if there are severe mixing issues. 

Finally, we hope that we have convinced readers that there are a number of very ingenious approaches and ideas to consider for tackling inference with intractable normalizing function problems. There are opportunities for  extending or even combining some of these algorithms -- it is not hard to believe that with some adjustments to the algorithms and clever uses of, say, parallel computing, any of these algorithms, for example the Russian roulette algorithm, is likely to become more efficient and hence more practical. Many of the ideas here relate to the even broader modern research problem of inference with models where likelihood functions are entirely unavailable or intractable. In the intractable normalizing function case it is apparent that there is still a dearth of generic, automated approaches for constructing efficient MCMC algorithms. For many models inference is infeasible even for moderately high-dimensional problems \citep[cf.][]{goldstein2014attraction}. Furthermore, careful approaches for comparing the efficiency and diagnosing convergence of asymptotically inexact algorithms is still an open problem. We hope that this manuscript will encourage further research on all of these challenging topics.


\clearpage

\appendix
\begin{center}
\title{\LARGE\bf Supplementary Material for Bayesian Inference in the Presence of Intractable Normalizing Functions}\\~\\

\author{\Large{Jaewoo Park and Murali Haran}}
\end{center}

The supplementary material provides details about how to select particles in the adaptive exchange (AEX) algorithm and a description of the Russian roulette algorithm. It also provides details about how computational complexity calculations for various algorithm was calculated. 

\section{Choice of particles in AEX}\label{particleapp}

\begin{algorithm}[hh]
\caption{Choosing particles in the AEX}\label{AEXparticle}
\begin{algorithmic}[H]
\normalsize

\State 1. Generate $\lbrace \theta_{1},...,\theta_{N_{0}} \rbrace$ from the fractional DMH.

\State 2. Choose $d$ particles from $N_{0}$ candidates through max-min procedure.

\State (a) Standardize each dimension of parameter to the interval $[0,1]$.

\State (b) Calculate distance matrix $[D_{st}]_{N_{0} \times N_{0}}$ of standardized parameters.

\State (c) Randomly choose one particle, $\theta^{(1)}=\theta_{k}$ among $N_{0}$ samples in the Step 1 and define $A=\lbrace k \rbrace$, $A^{c}=\lbrace 1,2,...,k-1,k+1,...,N_{0}\rbrace$. 

\State (d) Select the next particle $\theta^{(m)}$ as the farthest one, among the closest candidates from the previously chosen particles:\\
$\theta^{(m)}=\theta_{l}$ satisfying, $l \in A^{c}$ and $\exists ~ l' \in A$ such that $D_{l,l'}=\max_{s \in A^{c}} \min_{t \in A}  D_{st}$. Then set $A = A \cup \lbrace l \rbrace$ and $A^{c}=A^{c}\setminus\lbrace l \rbrace$. Repeat this for $m=2,...,d$.

\State 3. Return standardized particles in Step 2(a) to the original scale.
\end{algorithmic}
\end{algorithm}

~~~~\cite{liang2015adaptive} suggests an approach for choosing particles. Here, we provide the algorithm in detail , including information on how various algorithm components are to be tuned for choosing particles. In Step 1 of Algorithm~\ref{AEXparticle}, $N_{0}$ number of candidate particles are generated through fractional double Metropolis-Hastings (DMH). Fractional DMH is DMH with larger acceptance probability. By using an acceptance probability $\alpha^{\zeta}$ with $0<\zeta<1$ where 

\begin{equation}
\alpha = \min\left\lbrace 1, \frac{p(\theta')h(\mathbf{x}|\theta')h(\mathbf{y}|\theta_{n})q(\theta_{n}|\theta')}{p(\theta_{n})h(\mathbf{x}|\theta_{n})h(\mathbf{y}|\theta')q(\theta'|\theta_{n})}\right\rbrace,
\end{equation}

which is the acceptance probability of the double Metropolis-Hastings algorithm, we can draw samples from a wider region than the actual parameter space; samples can cover the support of the target posterior. Then a max-min procedure (see Steps 2 and 3 of Algorithm~\ref{AEXparticle}) is conducted to choose $d$ particles among $N_{0}$ candidates. Through this procedure, the $d$ selected particles have almost the same coverage in parameter space as $N_{0}$ candidates. We may use other strategies for choosing particles. For a social network model example, \cite{jin2013bayesian} selects particles through Approximate Bayesian Computation (ABC) \citep{beaumont2002approximate,marin2012approximate}, an approximate sampling scheme that does not involve evaluating the likelihood. 

\textbf{Components to be tuned}: To select particles, we have to decide the number of candidate particles ($N_{0}$), and the number of particles ($d$) in Algorithm~\ref{AEXparticle}. As the dimension of $\theta$ increases, we recommend using a larger $d$ to move around parameter space smoothly in the auxiliary chain. In this manuscript, we use $d=100$ for a one-dimension case, and $d=200$ for the four-dimension case. In both cases, a few thousand, $N_{0}=5,000$, appear to be adequate. 

\section{The Russian Roulette Algorithm }\label{Russianapp}

~~~~Let $\widetilde{Z}(\theta)$ be the upper bound of $Z(\theta)$ or any estimate of $Z(\theta)$. Then the likelihood is represented through multiplication with infinite geometric series as  

\begin{equation}
L(\theta|\mathbf{x})=\frac{h(\mathbf{x}|\theta)}{\widetilde{Z}(\theta)}c(\theta)\left[ 1+\sum_{n=1}^{\infty}k(\theta)^{n}\right],~~k(\theta)=1-c(\theta)\frac{Z(\theta)}{\widetilde{Z}(\theta)},
\label{e13}
\end{equation}

where $c(\theta)$ controls $|k(\theta)|<1$. Therefore with an infinite supply of independent unbiased estimates $\widehat{Z}_{i}(\theta)$ for $Z(\theta)$, an unbiased approximation of the likelihood can be constructed as follows.

\begin{equation}
\widehat{L}(\theta|\mathbf{x})=\frac{h(\mathbf{x}|\theta)}{\widetilde{Z}(\theta)}c(\theta)\left[ 1+\sum_{n=1}^{\infty}\prod_{i=1}^{n}\widehat{k}_{i}(\theta)\right],~~\widehat{k}_{i}(\theta)=1-c(\theta)\frac{\widehat{Z}_{i}(\theta)}{\widetilde{Z}(\theta)}.
\label{e14}
\end{equation}

Another possible choice is using multiplication correction with a Maclaurin series expansion of the exponential function \citep{lyne2015russian}. However both the geometric and Maclaurin series corrections require an infinite number of unbiased estimates $\widehat{Z}_{i}(\theta)$, which makes them impractical. To address this problem, \cite{lyne2015russian} propose stochastic truncation of the infinite series called the Russian Roulette, originally used in physics \citep{carter1975part}. Using simulated finite random stopping time $\tau$, only a $\tau$ number of $\widehat{Z}_{i}(\theta)$ are required to construct an unbiased likelihood approximation. This  leads to an asymptotically exact algorithm. 

Let the random stopping time be $\tau \geq 1$ with $p_{n}=P(\tau \geq n)$ for all $n \geq 1$, and $p_{1}=1$. The stopping time can be defined as $\tau=\inf\left\lbrace m\geq 1: U_{j} \geq q_{j} \right\rbrace$ where $U_{j} \sim$ i.i.d. $u(0,1), q_{j} \in (0,1]$, which makes $p_{n}=\prod_{j=1}^{n-1} q_{j}$. Then the partial sum $S_{\tau}= 1+\sum_{n=1}^{\tau-1}\prod_{i=1}^{n}\widehat{k}_{i}(\theta)/p_{n}$ with $S_{1}=1$ is called the Russian Roulette estimate. \cite{lyne2015russian} show that $S_{\tau}$ is the unbiased estimate of infinite series as  

\begin{equation}
E[S_{\tau}]= \sum_{j=1}^{\infty}S_{j}P(\tau=j) = 1+\sum_{n=1}^{\infty}\prod_{i=1}^{n}\widehat{k}_{i}(\theta).
\label{e18}
\end{equation}	

\begin{figure} 
\begin{center}
\includegraphics[scale=1.8]{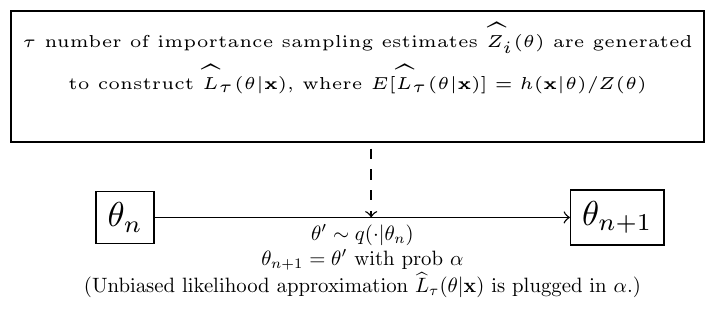}
\end{center}
\caption[]{Illustration for the Russian Roulette algorithm.}
\label{RussianFig}
\end{figure}

This indicates that the infinite series can be represented as the expectation of a finite series via the Russian Roulette truncation. In this way, an unbiased approximation of the likelihood is achieved by

\begin{equation}
\widehat{L}_{\tau}(\theta|\mathbf{x})=\frac{h(\mathbf{x}|\theta)}{\widetilde{Z}(\theta)}c(\theta)S_{\tau},~~S_{\tau}=\left[ 1+\sum_{n=1}^{\tau-1}\frac{\prod_{i=1}^{n}\widehat{k}_{i}(\theta)}{\prod_{j=1}^{n-1} q_{j}}\right],
\label{e19}
\end{equation}  

which makes the expectation of $\widehat{L}_{\tau}(\theta|\mathbf{x})$ equal \eqref{e14}. $\widehat{L}_{\tau}(\theta|\mathbf{x})$ will be plugged into the acceptance probability of the Metropolis-Hastings algorithm as in Figure~\ref{RussianFig}. This finite series truncation makes the algorithm practical, since the finite $\tau$ number of $\widehat{Z}_{i}(\theta)$ is used to construct unbiased approximation. The choice of $q_{j}$ directly has an impact on both variance of the likelihood approximation and the computing speed of the algorithm. Smaller $q_{j}$ can lead to smaller $\tau$ which is the required number of the $\widehat{Z}_{i}(\theta)$ for constructing $\widehat{L}_{\tau}(\theta|\mathbf{x})$. However \cite{lyne2015russian} provides a proof that if $q_{j}$ is close to zero, variance of likelihood approximation become large or even infinite. \cite{lyne2015russian} suggests that $q_{j}=\prod_{i=1}^{j}\widehat{k}_{i}(\theta)$ appears to work well in practice. 

Algorithm~\ref{RRalg} summarizes the Russian Roulette algorithm. Although the Russian Roulette algorithm is an asymptotically exact MCMC and applicable for general forms  of $h(\mathbf{x}|\theta)$, the algorithm is computationally expensive. It requires multiple ($\tau$) $\widehat{Z}_{i}(\theta)$s with each iteration (Step 3(b)). Each $\widehat{Z}_{i}(\theta)$ approximation in turn requires multiple MCMC samples from the likelihood, making it computationally expensive.

\begin{algorithm}
\caption{Russian Roulette algorithm}\label{RRalg}
\begin{algorithmic}[H]
\normalsize
\State Given $\theta_{n} \in \Theta$ at $n$th iteration.

\State 1. Propose $\theta' \sim q(\cdot|\theta_{n})$.

\State 2.  Calculate raw estimate $\widetilde{Z}(\theta')$ (e.g. importance sampling estimate).

\State 3. Construct unbiased likelihood estimate: start with $i=j=0$, $U_{0}=0,q_{0}=1$.

\While{$q_{j} > U_{j}$}\\

(a) $i=i+1$, $j=j+1$, and $U_{j} \sim$ i.i.d. $u(0,1)$.\\

(b) Construct a $\widehat{Z}_{i}(\theta')$ via importance sampling: 

$\widehat{Z}_{i}(\theta')=\frac{1}{N}\sum_{i=1}^{N} h(\mathbf{x}|\theta)/h(\mathbf{x}|\theta^{(0)})$. 

$\lbrace \mathbf{x}_{1},...,\mathbf{x}_{N} \rbrace \sim T^{m}_{\theta^{(0)}}(\cdot|\mathbf{x})$ whose stationary distribution is $h(\mathbf{x}|\theta^{(0)})/Z(\theta^{(0)})$, 

which is the importance function. \\

(c) Calculate $q_{j}=\prod_{i=1}^{j}\widehat{k}_{i}(\theta')$, where $\widehat{k}_{i}(\theta')=1-c(\theta')\widehat{Z}_{i}(\theta')/\widetilde{Z}(\theta')$.
\EndWhile

\State Set $\tau'=j$ and $\widehat{L}_{\tau'}(\theta'|\mathbf{x})=\frac{h(\mathbf{x}|\theta')}{\widetilde{Z}(\theta')}c(\theta')\left[ 1+\sum_{n=1}^{\tau'-1}\frac{\prod_{i=1}^{n}\widehat{k}_{i}(\theta')}{\prod_{j=1}^{n-1} q_{j}}\right]$.

\State 4. Accept $\theta_{n+1}=\theta'$ with probability 

$\alpha = \min\left\lbrace 1, \frac{p(\theta')\widehat{L}_{\tau'}(\theta'|\mathbf{x})q(\theta_{n}|\theta')}{p(\theta_{n})\widehat{L}_{\tau}(\theta_{n}|\mathbf{x})q(\theta'|\theta_{n})}\right\rbrace$, else reject (set $\lbrace \theta_{n+1},\mathbf{y}_{n+1}\rbrace=\lbrace \theta_n, \mathbf{y}_n \rbrace$).
\end{algorithmic}
\end{algorithm}

\textbf{Components to be tuned}: When a geometric correction is used for the Russian Roulette algorithm, five components of the algorithm need to be tuned: raw estimate $\widetilde{Z}(\theta)$, $c(\theta)$ which controls $|k(\theta)|<1$, ($N$) the number of samples for constructing the importance sampling estimate at each iteration, ($m$) length of inner sampler to generate MCMC samples, and the importance function $h(\mathbf{x}|\theta^{(0)})/Z(\theta^{(0)})$. If a tight upper bound of $Z(\theta)$ can be derived, it is a computationally efficient choice for $\widetilde{Z}(\theta)$, because it can lead to small $\widehat{k}_{i}(\theta)$ and small $q_{j}$, which implies small $\tau$. However, in most situation it is difficult to get a tight upper bound of $Z(\theta)$. Therefore, importance sampling approximations can be used for $\widetilde{Z}(\theta)$. $c(\theta)$ and $N$ should be determined through trial and error to make $|\widehat{k}_{i}(\theta)|<1$. In our simulation examples, we set $N$ between 1,000 and 2,000. $c(\theta)$=0.4 is used for every example. The length of the inner sampler $m$ should be larger as there is strong dependence among the data to generate more reliable samples. A simple choice for importance function is using maximum pseudolikelihood estimates as $\theta^{(0)}$. Other types of importance sampling estimate (e.g. annealed importance sampling) can also be used to construct $\widehat{Z}_{i}(\theta)$. To reduce computing time, $\widehat{Z}_{i}(\theta)$ may be obtained in parallel.

\textbf{Theoretical justification}: The Russian Roulette algorithm is based on the pseudo-marginal framework. If $h(\mathbf{x}|\theta)$ is bounded, we can attain a positive unbiased approximation of the likelihood and implement the algorithm without any theoretical difficulties. Even when $h(\mathbf{x}|\theta)$ is unbounded, $|\widehat{L}_{\tau}(\theta|\mathbf{x})|$ can be substituted instead of $\widehat{L}_{\tau}(\theta|\mathbf{x})$ in the Algorithm~\ref{RRalg} and weighted average of $f(\theta_{i})$ with sign of $\widehat{L}_{\tau}(\theta|\mathbf{x})$ is used to estimate $E[f(\theta)]$ without losing exactness. See "sign problem" in \cite{lyne2015russian} for details. Therefore, this algorithm is generally applicable.  

\subsection{Details of the simulation settings used in the examples}

~~~~\textbf{Ising model}: Simulation settings in the Russian
Roulette algorithm follow \cite{lyne2015russian}. When $\theta=0.2$, an annealed importance sampling (AIS) estimate of $Z(\theta)$ is constructed using 1,000 samples, and each sample is generated with 2,000 temperatures from an uniform lattice using the schedule. Geometric series correction with $c(\theta)=0.4$ is used to construct an unbiased estimate of likelihood. Same simulation settings are used for $\theta=0.43$ case except that 4,000 temperatures are used. Temperature schedules should be increased due to the slow mixing, which makes the algorithm more expensive. 

\textbf{Social network model}: 2,000 samples are generated through 20 cycles of Gibbs updates and used for importance sampling estimate whose importance function is $h(\mathbf{x}|\theta^{(0)})/Z(\theta^{(0)})$ where $\theta^{(0)}$ is the MPLE. A geometric series correction with $c(\theta)=0.4$ is used to construct an unbiased estimate of the likelihood. 

\textbf{Spatial point process}: 1,000 samples are generated through 2,000 iterations of birth-death MCMC and used for importance sampling approximations $\widehat{Z}_{i}(\theta)$ and $\widetilde{Z}(\theta)$. As an importance function, $h(\mathbf{x}|\theta^{(0)})/Z(\theta^{(0)})$ is used where $\theta^{(0)}$ is sample mean from short run of DMH. A geometric series correction with $c(\theta)=0.4$ is used for the Russian Roulette algorithm.

\section{Complexity calculations}\label{compapp}

~~~~~Here we provide details about calculating complexity and memory costs for the algorithms. Table~\ref{complexity} summarizes these calculations. In Table~\ref{complexity}, complexity indicates computational cost for a single run of the algorithms. Memory represents the magnitude of the quantities to be saved per each iteration to avoid re-computations. Fixed complexity measures all computational costs and fixed memory represents the size of variables to be saved before implementing the entire algorithm. We summarize our notation in Table~\ref{complexitynotation}. All other notations follow previous sections. 

\begin{table}[tt]
\centering
\begin{tabular}{cl}
  \hline
Symbol & Definition\\ 
  \hline
$n$ & the number of data points\\ 
$m$ & current iteration index in MCMC\\ 
$N$ & the number of samples used to construct importance sampling estimate\\ 
$N_{0}$ & the number of preliminary runs to attain particles (AEX, ALR)\\ 
$N_{1}$ & the number of preliminary runs for the auxiliary chain in AEX \\ 
$b$ & the thinning interval for MCMC\\ 
$c$ & the number of possible cores for parallel computing\\ 
$\tau$ & random stopping time (ALR, Russian Roulette)\\ 
$\widetilde{\tau}^{R}$ & expected stopping time for geometric Russian Roulette algorithm \\ 
$p$ & dimension of the parameter \\ 
$d$ & the number of particles \\
$S(\cdot)$ & sufficient statistics \\ 
$L(\cdot)$ & complexity function of evaluating $h(\cdot|\theta)$ \\ 
$G(\cdot)$ & complexity function of inner sampler \\
   \hline
\end{tabular}
\caption{Summary of notations used for complexity calculations.}
\label{complexitynotation}
\end{table}

\begin{table}[hh]
\centering
\begin{tabular}{ccc}
\hline
  DMH & Exponential family & Point process \\
  \hline
C & $G(n) + L(n)$ & $G(n^2) + 3L(n^2)$ \\ 
FM & $p$ & $n^2$ \\ 
    \hline
 NoisyDMH & Exponential family & Point process \\
  \hline
C & $N[G(n)+L(n)]/c $ & $NG(n^2)/c + [2N/c+1]L(n^2)$ \\ 
FM & $p$ & $n^2$ \\ 
\hline    
   $\ast$AEX & Exponential family & Point process \\
  \hline
C & $0.5[G(n)+L(n)]$ & $0.5G(n^2)+[N_{1}/b+m+5]L(n^2)$  \\ 
M & $p+2$ & $n^2+3$ \\ 
FC & $[N_{0}+0.5N_{1}][G(n)+L(n)]$ & $[N_{0}+0.5N_{1}]G(n^2)+[3N_{0}+N_{1}]L(n^2)$ \\
FM & $p+dp+d^2+(p+2)N_{1}/b$ & $n^2+dp+d^2+(n^2+3)N_{1}/b$ \\
    \hline    
   $\ast$ALR & Exponential family & Point process \\
  \hline
C & $G(n)+L(n)$ & $G(n^2)+ [(d + 2m+2)/c + 1]L(n^2)$ \\ 
M & $p+1$ & $n^2 + d+1$ \\ 
FC & $(N_{0}+\tau)[G(n)+L(n)]$  & $[N_{0}+\tau]G(n^2)+[3N_{0}+\tau d/c]L(n^2)$ \\
FM & $p+dp+d$ & $n^2+dp+d$ \\
    \hline    
 $\ast$Roulette & Exponential family & Point process \\
  \hline
C & $N(\widetilde{\tau}^{R} + 1)[G(n)+L(n)]/c$ & $N(\widetilde{\tau}^{R} + 1)G(n^2)/c + [2N(\widetilde{\tau}^{R} + 1)/c+1]L(n^2)$ \\ 
FM & $p$ & $n^2$ \\ 
    \hline
\end{tabular}
\caption{Computational complexity and memory size of the algorithms in the simulation examples. If costs are negligible, we didn't include in the table. C, M, FC, FM denote complexity, memory, fixed complexity, fixed memory respectively. Asterisk symbols indicate calculated complexity is expected number.}
\label{complexity}
\end{table}

\subsection{Complexity of DMH}
In the double Metropolis-Hastings algorithm
\begin{enumerate}
\item Exponential family
\begin{itemize}
\item Complexity\\
Step 2: A single auxiliary variable $\mathbf{y}$ is generated.\\ 
Step 3: $S(\mathbf{y})$ is calculated once.\\ 
\textbf{Total}: $G(n)+L(n)$
\item Fixed memory\\
Data: $p$-dimensional $S(\mathbf{x})$ should be stored.\\
\textbf{Total}: $p$
\end{itemize}
\item Point process 
\begin{itemize}
\item Complexity\\
Step 2: A single auxiliary variable $\mathbf{y}$ is generated.\\ 
Step 3: $h(\mathbf{x}|\theta'),h(\mathbf{y}|\theta),h(\mathbf{y}|\theta')$ are calculated.\\ 
\textbf{Total}: $G(n^2)+3L(n^2)$
\item Fixed memory\\
Data: $n$ by $n$ distance matrix of $\mathbf{x}$ should be stored.\\
\textbf{Total}: $n^2$
\end{itemize}
\end{enumerate}

\subsection{Complexity of Noisy DMH}	
In the noisy double Metropolis-Hastings algorithm
\begin{enumerate}
\item Exponential family
\begin{itemize}
\item Complexity\\
Step 2: $N$ number of auxiliary variables $\mathbf{y}_{j}$ are generated.\\ 
Step 3: $N$ number of corresponding $S(\mathbf{y}_{j})$ are calculated.\\ 
\textbf{Total}: Steps can be parallelized, $N[G(n)+L(n)]/c$
\item Fixed memory\\
Data: $p$-dimensional $S(\mathbf{x})$ should be stored.\\
\textbf{Total}: $p$
\end{itemize}
\item Point process 
\begin{itemize}
\item Complexity\\
Step 2: $N$ number of auxiliary variables $\mathbf{y}_{j}$ are generated.\\ 
Step 3: $N$ number of $h(\mathbf{y}_{j}|\theta),h(\mathbf{y}_{j}|\theta')$ and a single $h(\mathbf{x}|\theta')$ are calculated.\\ 
\textbf{Total}: Generating $N$ number of $\mathbf{y}_{j}$ and evaluating $N$ number of $h(\mathbf{y}_{j}|\theta),h(\mathbf{y}_{j}|\theta')$ can be parallelized, $N[G(n^2)+2L(n^2)]/c + L(n^2)$
\item Fixed memory\\
Data: $n$ by $n$ distance matrix of $\mathbf{x}$ should be stored.\\
\textbf{Total}: $n^2$.
\end{itemize}
\end{enumerate}

\subsection{Complexity of AEX}
In the adaptive exchange algorithm
\begin{enumerate}
\item Exponential family
\begin{itemize}
\item Complexity\\
Step 1(b): With $0.5$ probability, one MH update of $\mathbf{x}'$ and $S(\mathbf{x}')$ is calculated once.\\
\textbf{Total}: $0.5[G(n)+L(n)]$
\item Memory\\
$p$-dimensional $S(\mathbf{x}_{m+1})$ and scalar $w_{m+1}^{(I_{m+1})}$, $I_{m+1}$ are stored.\\ 
\textbf{Total}: $p+2$
\item Fixed complexity\\
Choosing particles: $N_{0}$ iterations of fractional DMH, $N_{0}[G(n)+L(n)]$\\
Preliminary run for auxiliary chain: $0.5N_{1}[G(n)+L(n)]$\\ 
\textbf{Total}: $[N_{0}+0.5N_{1}][G(n)+L(n)]$
\item Fixed memory\\
Data: $p$-dimensional $S(\mathbf{x})$ should be stored.\\
Particles: $d$ number of $p$-dimensional particles, and $d$ by $d$ distance matrix of particles to define neighbors should be stored.\\
Preliminary run for auxiliary chain: $p$-dimensional $S(\mathbf{x}_{m+1})$ and scalar $w_{m+1}^{(I_{m+1})}$, $I_{m+1}$ are stored for $N_{1}$ iterations, and can be stored with $b$ thinning interval.\\
\textbf{Total}: $p+dp+d^2+(p+2)N_{1}/b$
\end{itemize}

\item Point process 
\begin{itemize}
\item Complexity\\
Step 1(a): With $0.5$ probability, $h(\mathbf{x}_{m}|\theta^{(I')})$ is calculated once.\\ 
Step 1(b): With $0.5$ probability, one MH update of $\mathbf{x}'$ and  $h(\mathbf{x}'|\theta^{(I_{m})})$ is calculated once.\\
Step 5: $|H_{m+1}|$ number of $h(\mathbf{x}_{j}|\theta')$ should be calculated. With $N_{1}$ number of preliminary run and $b$ thinning interval for the auxiliary chain, $|H_{m+1}|$ is equal to $N_{1}/b+m+1$. \\
Step 6: $h(\mathbf{x}|\theta'),h(\mathbf{y}|\theta),h(\mathbf{y}|\theta')$ are calculated.\\ 
\textbf{Total}: $0.5G(n^2)+[N_{1}/b+m+5]L(n^2)$
\item Memory\\
$n$ by $n$ distance matrix of $\mathbf{x}_{m+1}$, scalar $w_{m+1}^{(I_{m+1})}$, $I_{m+1}$ and evaluated $h(\mathbf{x}_{m+1}|\theta^{(I_{m+1})})$ are stored.\\
\textbf{Total}: $n^2+3$
\item Fixed complexity\\
Choosing particles: $N_{0}$ iterations of fractional DMH, $N_{0}[G(n^2)+3L(n^2)]$\\
Preliminary run for auxiliary chain: $N_{1}[0.5L(n^2)+0.5[G(n^2)+L(n^2)]]$\\
\textbf{Total}: $[N_{0}+0.5N_{1}]G(n^2)+[3N_{0}+N_{1}]L(n^2)$
\item Fixed memory\\
Data: $n$ by $n$ distance matrix of $\mathbf{x}$ should be stored.\\
Particles: $d$ number of $p$-dimensional particles, and $d$ by $d$ distance matrix of particles to define neighbors should be stored.\\ 
Preliminary run for auxiliary chain: $n$ by $n$ distance matrix of $\mathbf{x}_{m+1}$, scalar $w_{m+1}^{(I_{m+1})}$, $I_{m+1}$ and evaluated $h(\mathbf{x}_{m+1}|\theta^{(I_{m+1})})$ are stored for $N_{1}$ iterations, and can be stored with $b$ thinning interval.\\ 
\textbf{Total}: $n^2+dp+d^2+(n^2+3)N_{1}/b$
\end{itemize}
\end{enumerate}

In practice, AEX consists of three parts: (1) choosing particles, (2) preliminary run for the auxiliary chain, and (3) implementing the entire algorithm. Fixed complexity and fixed memory are evaluated from the first two parts.  
   
\subsection{Complexity of ALR}
In the ALR algorithm
\begin{enumerate}
\item Exponential family
\begin{itemize}
\item Complexity\\
Step 1: $\mathbf{x}_{m+1}$ is generated once.\\
Step 2: $S(\mathbf{x}_{m+1})$ is calculated once.\\ 
\textbf{Total}: $G(n)+L(n)$
\item Memory\\
$p$-dimensional $S(\mathbf{x}_{m+1})$ and scalar $I_{m+1}$ is stored.\\
\textbf{Total}: $p+1$
\item Fixed complexity\\
Choosing particles: $N_{0}$ iterations of DMH, $N_{0}[G(n)+L(n)]$\\
Stopping time: Step 1-3 are repeated until $\tau$, $\tau[G(n)+L(n)]$\\
\textbf{Total}: $(N_{0}+\tau)[G(n)+L(n)]$
\item Fixed memory\\
Data: $p$-dimensional $S(\mathbf{x})$ should be stored.\\
Particles: $d$ number of $p$-dimensional particles are stored.\\
Stopping time: $d$-dimensional $\mathbf{c}_{m}$ is stored.\\
\textbf{Total}: $p+dp+d$
\end{itemize}

\item Point process
\begin{itemize}
\item Complexity\\
Step 1: $\mathbf{x}_{m+1}$ is generated once.\\
Step 2: $d$ number of $h(\mathbf{x}_{m+1}|\theta^{(i)})$ are calculated, which can be parallelized.\\ 
Step 4: $h(\mathbf{x}_{j}|\theta'),h(\mathbf{x}_{j}|\theta)$ are evaluated for $m+1$ number of cumulative $\mathbf{x}_{j}$, which can be parallelized.\\
Step 5: $h(\mathbf{x}|\theta')$ is evaluated once.\\
\textbf{Total}: $G(n^2)+[(d+2m+2)/c+1]L(n^2)$
\item Memory\\
$n$ by $n$ distance matrix of $\mathbf{x}_{m+1}$, scalar $I_{m+1}$ and $d$ number of $h(\mathbf{x}_{m+1}|\theta^{(i)})$ should be stored.\\
\textbf{Total}: $n^2+d+1$
\item Fixed complexity\\
Choosing particles: $N_{0}$ iterations of DMH, $N_{0}[G(n^2)+3L(n^2)]$\\
Stopping time: Step 1-3 are repeated until $\tau$, $\tau[G(n^2)+dL(n^2)/c]$\\
\textbf{Total}: $[N_{0}+\tau]G(n^2)+[3N_{0}+\tau d/c]L(n^2)$
\item Fixed memory\\
Data: $n$ by $n$ distance matrix of $\mathbf{x}$ should be stored.\\
Particles: $d$ number of $p$-dimensional particles are stored.\\
Stopping time: $d$-dimensional $\mathbf{c}_{m}$ is stored.\\ 
\textbf{Total}: $n^2+dp+d$
\end{itemize}

\end{enumerate}

Similar to AEX, the algorithm consists of three parts: (1) choosing particles, (2) updating $\lbrace \mathbf{x}_{m}, I_{m}, \mathbf{c}_{m} \rbrace$, until $\mathbf{c}_{m}$ converges, and (3) updating the entire process. Fixed complexity and fixed memory are evaluated from the first two parts.  

It is observed that random stopping time $\tau$ is affected by the complexity of the parameter space and choice of particles rather than $n$. Intuitively, as the particles cover important regions of the parameter space, it is easier to move around the state space quickly. Therefore, proper choice of particles can lead to smaller $\tau$.  We observe that if we choose particles covering interesting region of parameter space as in Algorithm~\ref{AEXparticle}, $\tau$ is almost fixed even with increasing $n$.

\subsection{Complexity of the Russian Roulette algorithm}

In the Russian roulette algorithm
\begin{enumerate}
\item Exponential family
\begin{itemize}
\item Complexity\\
Step 2: Importance sampling estimate, $\widetilde{Z}(\theta')$ is constructed, which requires generating $N$ number of $\mathbf{y}_{j}$ and calculating $S(\mathbf{y}_{j})$.\\  
Step 3: $\tau$ number of $\widehat{Z}_{i}(\theta')$ are constructed, which require generating $N$ number of $\mathbf{y}_{j}$ and calculating $S(\mathbf{y}_{j})$ per each $\widehat{Z}_{i}(\theta')$.\\  
\textbf{Total}: Steps can be parallelized, $N(\tau+1)[G(n)+L(n)]/c$
\item Fixed Memory\\
Data: $p$-dimensional $S(\mathbf{x})$ should be stored.\\
\textbf{Total}: $p$
\end{itemize}
\item Point process 
\begin{itemize}
\item Complexity\\
Step 2: Importance sampling estimate, $\widetilde{Z}(\theta')$ is constructed, which requires generating $N$ number of $\mathbf{y}_{j}$ and calculating $h(\mathbf{y}_{j}|\theta'),h(\mathbf{y}_{j}|\theta^{(0)})$.\\ 
Step 3: $\tau$ number of $\widehat{Z}_{i}(\theta')$ are constructed, which require generating $N$ number of $\mathbf{y}_{j}$ and calculating $h(\mathbf{y}_{j}|\theta'),h(\mathbf{y}_{j}|\theta^{(0)})$ per each $\widehat{Z}_{i}(\theta')$. Then $h(\mathbf{x}|\theta')$ is calculated.\\  
\textbf{Total}: Step 2-3 can be parallelized, $N(\tau+1)[G(n^2)+2L(n^2)]/c + L(n^2)$
\item Fixed Memory\\
Data: $n$ by $n$ distance matrix of $\mathbf{x}$ should be stored.\\
\textbf{Total}: $n^2$.
\end{itemize}
\end{enumerate}

\begin{equation}
E[\tau]=\sum_{x=1}^{\infty}xp(\tau=x) = \sum_{x=1}^{\infty}xk^{x(x-1)/2}(1-k^{x}) = \sum_{x=1}^{\infty}k^{x(x-1)/2}=\widetilde{\tau}^{R} 
\label{e28}
\end{equation}

Because the stopping time $\tau$ is a random variable, we propose an approximate evaluation of the expected stopping time. Consider $q_{j}$, where $p(\tau \geq x)=\prod_{j=1}^{x-1}q_{j}$ as we defined in the previous section. We can use $q_{j}=\prod_{i=1}^{j}\widehat{k}_{i}(\theta)$, where $\widehat{k}_{i}(\theta)=1-c(\theta)\widehat{Z}_{i}(\theta)/\widetilde{Z}(\theta)$ as suggested in \cite{lyne2015russian}. For large enough  importance sampling size $N$, $\widehat{Z}(\theta)_{i}/\widetilde{Z}(\theta)$ will be close to 1 when $\widetilde{Z}(\theta)$ is importance sampling estimate as well. In this case, $\widehat{k}_{i}(\theta) \approx k=0.6$  is almost constant because $c(\theta)=0.4$ is used for our examples. Therefore, $q_{j} \approx k^{j}$ and expected stopping time can be represented as \eqref{e28}. For $c(\theta)=0.4$, $\widetilde{\tau}^{R} \approx 1.86$.   

Then remaining issue is how large $N$ should be depending on size of data $n$. $N$ should be tuned to make $|\widehat{k}_{i}(\theta)|<1$ for convergence of geometric series. As $N$ increases, it is oberved that the mixing of the chain become better (higher effective sample size) and ergodic average with lower Monte Carlo standard error can be attained at the expense of increased computational costs. How to tune $N$ efficiently is still debatable. In pseudo-marginal MCMC, \cite{doucet2015efficient} propose the rule of thumb to choose $N$ so that standard deviation of the log-likelihood estimate is around 1.2. Although this can be a one guideline to tune $N$, it is found that suggested $N$ is too large in our point process example which can lead unnecessary costs.  Also, it is found that using Annealed importance sampling estimate can achieve lower variance in many problems than using simple importance sampling estimate. Considering all possibilities, it is challenging to generalize how $N$ should be tuned depending on $n$, because it can be problem specific.

\bibliography{Reference}

\end{document}